\documentclass[11pt,a4paper]{article}

\pdfoutput=1
\usepackage{jheppub}

\usepackage{braket}

\usepackage{graphicx}
\usepackage{wrapfig,enumerate,slashed}

\usepackage{footmisc}
\usepackage{amsmath}
\usepackage{wasysym} 
\usepackage{color}
\usepackage{multirow}

\DeclareMathAlphabet{\mathbold}{U}{zeur}{b}{n}

\usepackage[labelfont=bf,font={sl,small}]{caption}
\newcommand{\mc}[1]{#1}

\renewcommand\[{\left[}
\renewcommand\]{\right]}

\def\beq{\begin{equation}}
\def\eeq{\end{equation}}
\def\[{\begin{equation}}
\def\]{\end{equation}}

\begin{document}
\numberwithin{equation}{section}

\title{Searches for vector-like quarks at future\\[0.2cm] colliders and implications for composite Higgs models \\[0.2cm]  with dark matter}

\author{Mikael Chala,}
\author{Ramona Gr\"ober}
\author{and Michael Spannowsky}

\affiliation{Institute for Particle Physics Phenomenology, Department of Physics, Durham University, Durham, DH1 3LE, UK}

\emailAdd{mikael.chala@durham.ac.uk}
\emailAdd{ramona.groeber@durham.ac.uk}
\emailAdd{michael.spannowsky@durham.ac.uk}

\abstract{Many composite Higgs models predict the existence of vector-like quarks with masses outside the reach of the LHC, \textit{e.g.} $m_Q \gtrsim 2$ TeV, in particular if these models contain a dark matter candidate. In such models the mass of the new resonances is bounded from above to satisfy the constraint from the observed relic density. We therefore develop new strategies to search for vector-like quarks at a future $100$ TeV collider and evaluate what masses and interactions can be probed. We find that masses as large as $\sim 6.4$ ($\sim 9$) TeV can be tested if the fermionic resonances decay into Standard Model (dark matter) particles. We also discuss the complementarity of dark matter searches, showing that most of the parameter space can be closed. On balance, this study motivates further the consideration of a higher-energy hadron collider for a next generation of facilities.
}

\preprint{IPPP/18/6}

\maketitle

\section{Introduction}
Composite Higgs Models (CHMs)~\cite{Dimopoulos:1981xc,Kaplan:1983fs,Kaplan:1983sm} are among the most compelling solutions to the hierarchy problem. They involve a new strongly-interacting sector (approximately) symmetric under a global group $\mathcal{G}$ that is spontaneously broken to $\mathcal{H}\subset\mathcal{G}$ at a new physics scale $f$. The Higgs boson is assumed to be a pseudo Nambu-Goldstone Boson (pNGB) of this symmetry breaking pattern. The smoking-gun signature of this setup is the presence of new resonances, in particular Vector-Like Quarks (VLQs), whose masses scale as $m_\rho\sim g_\rho f$, with $g_\rho$ being the coupling of the strong sector. The Higgs mass in these models is also expected to scale as \mc{$\sim m_\rho/(4\pi)$}, hence requiring some tuning to make it light ($\sim 125$ GeV) for large $m_\rho$.

Current experimental bounds on $m_\rho$ range from $\sim 900$ to $\sim 1300$ GeV for top-like VLQs\footnote{We emphasize that the corresponding searches are motivated by the minimal CHM~\cite{Agashe:2004rs}. Thus, the results are obtained under the assumption that VLQs decay only into SM particles. For non-minimal symmetry breaking, the bounds can be significantly altered~\cite{Chala:2017xgc}.}~\cite{Araque:2016jrb,Aaboud:2017zfn,Aaboud:2017qpr}. 
These numbers are still compatible with expectations from naturalness arguments: \textit{e.g.} Refs.~\cite{Contino:2006qr,Matsedonskyi:2012ym,Redi:2012ha,Marzocca:2012zn,Pomarol:2012qf,Panico:2012uw} highlighted values for $m_\rho$ in the range $\lesssim 1-1.5$ TeV. Moreover, latter references (\textit{e.g.} Ref.~\cite{Panico:2012uw}) have also shown that masses as large as $m_\rho \gtrsim 2$ TeV are compatible with a tuning on the Higgs mass of the order of $\sim 1/100$ in some classes of CHMs.
Such masses are most probably beyond LHC reach~\cite{Azatov:2013hya, Ortiz:2014iza, Matsedonskyi:2015dns} (at least in the model-independent pair-production mode). Therefore, it is conceivable that VLQs might not be discovered at the LHC and new facilities will be  required to probe such models.

This conclusion can be further strengthened by exploring non-minimal CHMs containing extra stable pNGBs that can play the role of Dark Matter (DM) particles. These models are well motivated by two main reasons. \textit{(i)} One single mechanism explains why the electroweak (EW) and the DM scales are of the same order, as suggested by the WIMP paradigm. \textit{(ii)} The Higgs boson can have naturally small portal couplings to the pNGB DM, which evades the strong constraints from low-energy direct detection experiments. At the same time, the observed relic density can be produced by effective derivative couplings $\sim 1/f^2 (\partial S)^2 h^2$ ($\sim m_S^2/f^2$ at momenta $p\sim m_S$ at which DM annihilation occurs). Thus, we will show that scenarios of this kind require $m_\rho\sim$ few TeV to accommodate the correct DM abundance. Consequently, we study the reach of a future $100$ TeV collider to the VLQs, including decays into SM particles, which are also relevant for the minimal CHM, as well as into DM particles.

We extend previous works on the interplay between collider and DM searches in CHMs~\cite{Frigerio:2012uc, Marzocca:2014msa, Fonseca:2015gva} in several ways: \textit{(i)} Instead of focusing on a particular model, we adopt a generic parametrization that captures the main features of cosets like $SO(6)/SO(5)$~\cite{Gripaios:2009pe}, $SO(7)/SO(6)$~\cite{Chala:2016ykx,Balkin:2017aep}, $SO(7)/G_2$~\cite{Chala:2012af,Ballesteros:2017xeg}, $SO(5)\times U(1)/SO(4)$~\cite{Gripaios:2016mmi}, etc. \textit{(ii)} We match to representations not previously considered in the literature (\textit{e.g} the $20$ in $SO(6)/SO(5)$) as well as symmetry breaking patterns not yet studied (\textit{e.g.} $SO(7)/SO(5)$ or $SO(6)/SO(4)$). \textit{(iii)} In what concerns LHC constraints, we consider the latest experimental data, including LHC searches for heavy pair-produced resonances at $13$ TeV. \textit{(iv)} We quantify the effect of having all resonances of a multiplet at once, instead of considering constraints on each separately. 

This article is organized as follows. In Section~\ref{sec:parameterization}, we introduce briefly the effective parametrization describing the different models of interest. In Section~\ref{sec:dm} we discuss the DM phenomenology in light of the parameters discussed previously. In Section~\ref{sec:collider} we detail the current LHC constraints on new fermionic resonances and discuss new analysis strategies for future colliders. We devote Section~\ref{sec:vlqs} to searches for VLQs with SM decays and Section~\ref{sec:stops} to searches for VLQs decaying into DM particles. In Section~\ref{sec:matching} we match the coefficients of the aforementioned parametrization to concrete non-minimal CHMs. We discuss the interplay between collider and DM searches in Section~\ref{sec:interplay} and highlight the characteristics of the most viable models. 
\section{Parametrization}\label{sec:parameterization}
We will denote by $H$ the SM Higgs doublet with hypercharge $ Y = 1/2$. Likewise, we assume the presence of a single scalar DM field $S$, singlet under the SM gauge group, whereas odd under a $\mathbb{Z}_2$ symmetry $S\rightarrow -S$. The relevant Lagrangian for our study is parameterized by $m_\rho$ and $g_\rho$, namely the typical mass of the fermionic resonances and the typical coupling of the strong sector, as well as a number of dimensionless coefficients. We can write the Lagrangian explicitly as 
\begin{align}\label{eq:parameterization}\nonumber
 L &= |D_\mu H|^2 \left[1-a_1\frac{S^2}{f^2}\right]  + \frac{a_2}{f^2}\partial_\mu |H|^2 (S\partial_\mu S)+ \frac{1}{2}(\partial_\mu S)^2 \left[1-2a_3\frac{|H|^2}{f^2}\right]\\
 &-m_\rho^2 f^2 \frac{N_c y_t^2}{(4\pi)^2} \left[-\alpha \frac{|H|^2}{f^2} + \beta\frac{|H|^4}{f^4} + \gamma\frac{S^2}{f^2} + \delta \frac{S^2 |H|^2}{f^4}\right] + \left[i\epsilon \frac{y_t}{f^2} S^2 \overline{q}_L H t_R+\text{h.c.}\right]~,
\end{align}
where $a_1, a_2, a_3$ and $\alpha,\beta,\gamma,\delta, \epsilon$ are $\mathcal{O}(1)$, $f=m_\rho/g_\rho$, $N_c$ is the number of $SU(3)$ colours and $y_t\sim 1$ is the top Yukawa coupling. Note that not all these parameters are physical: for instance, a scaling of $f$ could be reabsorbed in the dimensionless coefficients. Likewise, only a particular combination of these coefficients enter into physical observables, \textit{e.g.} the DM annihilation cross section (see below). This parametrization is simple, predictive, yet flexible enough, and it can comprise very different CHMs. Moreover, it reflects the expected power counting in these setups~\cite{Giudice:2007fh,Chala:2017sjk}. Finally, we emphasize that, being a strong coupling, $g_\rho$ is expected to be $\gtrsim 1$, while perturbative unitarity implies\footnote{The $\sqrt{2\pi}$ reduction on this estimation in comparison with the naive $4\pi$ has been pointed out \textit{e.g.} in Refs.~\cite{Lee:1977eg,Panico:2015jxa}. Often, the upper value $\sqrt{4\pi}$ is also used in the literature.} $g_\rho\lesssim 2\sqrt{2\pi}\sim 5$. We restrict ourselves to this range henceforth.

One can easily link the phenomenology of pair-produced VLQs with that of $S$. As a matter of fact, the former depends only on the mass of the VLQs, which is just given by $m_\rho \sim g_\rho f$. The number of such resonances and their charges depend crucially on the coset structure. In all our cases of interest, however, there is always a fourplet of VLQs transforming as $(\mathbf{2}, \mathbf{2})$ under the custodial group $SU(2)_L\times SU(2)_R$ and/or a VLQ decaying 100\% into $St$, with $h$ the physical Higgs boson and $t$ the SM top quark. For concreteness we assume that the decay rates of the different components in the fourplet are~\cite{Serra:2015xfa}
\begin{align}\label{eq:partners}\nonumber
 &\text{BR}(T, X_{2/3}\rightarrow ht) \sim \text{BR}(T, X_{2/3}\rightarrow Zt)\sim 0.5~,\\
 &\text{BR}(B\rightarrow W^-t)\sim \text{BR}(X_{5/3}\rightarrow W^+t)\sim \text{BR}(T^\prime\rightarrow St)\sim 1~.
\end{align}

\section{Dark matter phenomenology}\label{sec:dm}
Contrary to the derivative interactions in Eq.~\ref{eq:parameterization}, the effective coupling driven by $\epsilon$ is not enhanced by the DM mass, $m_S$, at the annihilation scale. Additionally, it is suppressed by an additional $1/f$ factor with respect to the Higgs portal coupling in the potential (proportional to $\delta$) in the low-energy DM-nucleon interactions. For these reasons, we take it to be zero hereafter for simplicity, \mc{but we will comment on the implications of switching it on when relevant}.

Thus, the annihilation cross section is driven by the $|H|^2 S^2$ interaction, receiving contributions from both the sigma model Lagrangian and the potential. In particular, the Feynman rule associated to the quartic coupling between two DM particles and two Higgses (or two longitudinal gauge bosons) reads
\begin{align}\label{eq:vertex}\nonumber
V &= \frac{i}{f^2}\left[-2N_c\delta\frac{m_\rho^2 }{(4\pi)^2} + 2a_1(p_1\cdot p_2) + 2a_3(p_3\cdot p_4) - a_2(p_1+p_2)(p_3+p_4) \right]~\\
&=\frac{2i}{f^2}\left[(2a_1 + 2a_2 + a_3)m_S^2 - N_c\delta\frac{m_\rho^2 }{(4\pi)^2}\right]\sim \frac{2i N_c m_\rho^2}{(4\pi)^2f^2}\left[2(2a_1 + 2a_2 + a_3)\gamma - \delta\right]~,
\end{align}
where
$p_1$ and $p_2$ are the four-momenta of the Goldstone bosons while $p_3$ and $p_4$ stand for those of the DM particles.  All momenta are assumed to be incoming. Due to momentum conservation we have the relation $(p_1+p_2) = -(p_3+p_4)$. The last equality in Eq.~(\ref{eq:vertex}) holds in the limit in which $m_S \gg m_h, m_W, m_Z$.
We have neglected $\mathcal{O}(v^2/f^2)$ terms in $m_S$, where $v\sim 246$ GeV denotes the EW vacuum expectation value.
In CHMs where both $S$ and $H$ come in the same multiplet of a larger group $\mathcal{G}$, one expects $a_1=a_2=a_3 \equiv a$. In such a case, the derivative interactions drive the DM annihilation provided $ a\gamma > \delta/10$. Given that all these couplings are expected to be $\lesssim\mathcal{O}(1)$, derivative interactions are expected to be highly relevant in this class of models. Similar results have been previously pointed out in Refs.~\cite{Frigerio:2012uc,Fonseca:2015gva,Chala:2016ykx,Bruggisser:2016ixa,Ballesteros:2017xeg,Balkin:2017aep,Balkin:2017yns}. \mc{For $a\gamma \sim \delta/10$, $V$ in Eq.~\ref{eq:vertex} is very small, and therefore $\epsilon \neq 0$ can dominate the DM annihilation rate; see Ref.~\cite{Balkin:2017yns} for an explicit example.}

Regarding $g_\rho$ and $f$, the relic density scales as $\Omega h^2 \sim m_S^2/V^2 \sim f^2/g_\rho^2$.
As a consequence, very large values of $f$, as well as very small values of $g_\rho$, are excluded by the requirement $\Omega h^2 \leqslant \Omega h^2_{\text{obs}} \sim 0.11$, where $\Omega h^2_{\text{obs}}$ stands for the measured value of the total relic abundance~\cite{Ade:2013zuv}.
Concerning direct detection constraints, derivative interactions are irrelevant as they are velocity suppressed. We consider current LUX limits~\cite{Akerib:2016vxi} and future LZ goals~\cite{Szydagis:2016few} on spin-independent cross sections. In our notation, the theoretical prediction for the spin-independent cross section reads
\begin{equation}
 \sigma \sim \frac{9}{256 \pi^5}  m_N^2 f_N^2 \delta^2 \frac{ g_\rho^4 }{ m_h^4 m_S^2}~,
\end{equation}
with $f_N\sim 0.3$~\cite{Alarcon:2011zs,Alarcon:2012nr} and $m_N\sim 1$ GeV. Bounds from direct searches are therefore complementary to those set by the upper limit on $\Omega h^2$. Furthermore, both are parametrically complementary to the quantity $g_\rho f$ tested by collider searches for VLQs. Finally, indirect searches are of little relevance for scalar singlet models like the ones considered here~\cite{Ballesteros:2017xeg,Duerr:2015aka}.

\section{Searches for new resonances}\label{sec:collider}
The masses of the top partners in Eq.~\ref{eq:partners} scale like $m_\rho = g_\rho f$. Thus, collider searches for VLQs are also complementary to the bounds set by direct-detection experiments and the measurement of the relic abundance. Moreover, contrary to direct detection tests, they are independent of the value of $\delta$. 

In order to compute the reach of current LHC data to the heavy resonances, we use \texttt{VLQlimits}~\cite{Chala:2017xgc}. This code includes the information of several experimental searches, at both 8 TeV~\cite{Aad:2015kqa} and 13 TeV~\cite{ATLAS-CONF-2016-102,ATLAS-CONF-2016-104,ATLAS-CONF-2017-015} with the largest luminosity, as well as SUSY searches sensitive to  $T'\to St$~\cite{CMS:2016hxa}. The code takes into account the simultaneous presence of all vector-like fermions in Eq.~\ref{eq:partners}. The limits obtained in this way set a robust constraint on 
\begin{equation}
 m_\rho = g_\rho f < 1.2~\text{TeV}~.
\end{equation}
For a high-luminosity run of the LHC with $\mathcal{L} = 3$ ab$^{-1}$, we estimate the corresponding bound by rescaling the signal and background events with the luminosity. We obtain
\begin{equation}
 m_\rho = g_\rho f < 1.7~\text{TeV}~.
\end{equation}
\subsection{Search for vector-like quarks with Standard Model decays at 100 TeV}\label{sec:vlqs}
In order to estimate the prospects for the searches of VLQs at 100 TeV, we consider a 3-lepton final state. In case of $2/3$ charged VLQs such a final state arises in $T \to Zt$ decays where one $Z$ boson and one top quark decays leptonically. For $5/3$ charged or $-1/3$ charged VLQs this final state arises from decays $X\to Wt$. While at the LHC searches for VLQs are mainly performed in 1-or 2-lepton final states (with the exception of the search \cite{Aad:2014efa}), due to the larger cross section, the 3-lepton final state has the advantage of smaller backgrounds. Since the cross sections are in general larger at the 100 TeV collider than at the LHC, the 3-lepton final state is an optimal choice for an estimate of the prospects for searches of VLQs.

We generate the signal with {\tt Madgraph5\_aMC@NLO} \cite{Alwall:2014hca} and use {\tt HERWIG} \cite{Bellm:2015jjp,Bellm:2017bvx} for the parton shower and the hadronization. The model file for the signal was generated with {\tt FeynRules}~\cite{Alloul:2013bka}. The dominant background processes are $t\bar{t}VV$, $t\bar{t}t\bar{t}$ and $t\bar{t}V+\text{jets}$, with $V = W^\pm, Z$. For the latter we generate samples with an exclusive jet and we merge an additional jet  using {\tt Sherpa 2.2.2} \cite{Gleisberg:2008ta, Gleisberg:2008fv}. The background events for the processes $t\bar{t}VV$ and $t\bar{t}t\bar{t}$ are generated with {\tt Madgraph5\_aMC@NLO} and showered and hadronized with {\tt HERWIG}. Jets are clustered with the anti-$k_t$ algorithm as implemented in {\tt FastJet} \cite{Cacciari:2011ma}, with a radius parameter $R=0.4$. For the analysis we use {\tt Rivet} \cite{Buckley:2010ar}.\\

\begin{figure}[t]
\begin{center}
\includegraphics[width=0.6\columnwidth]{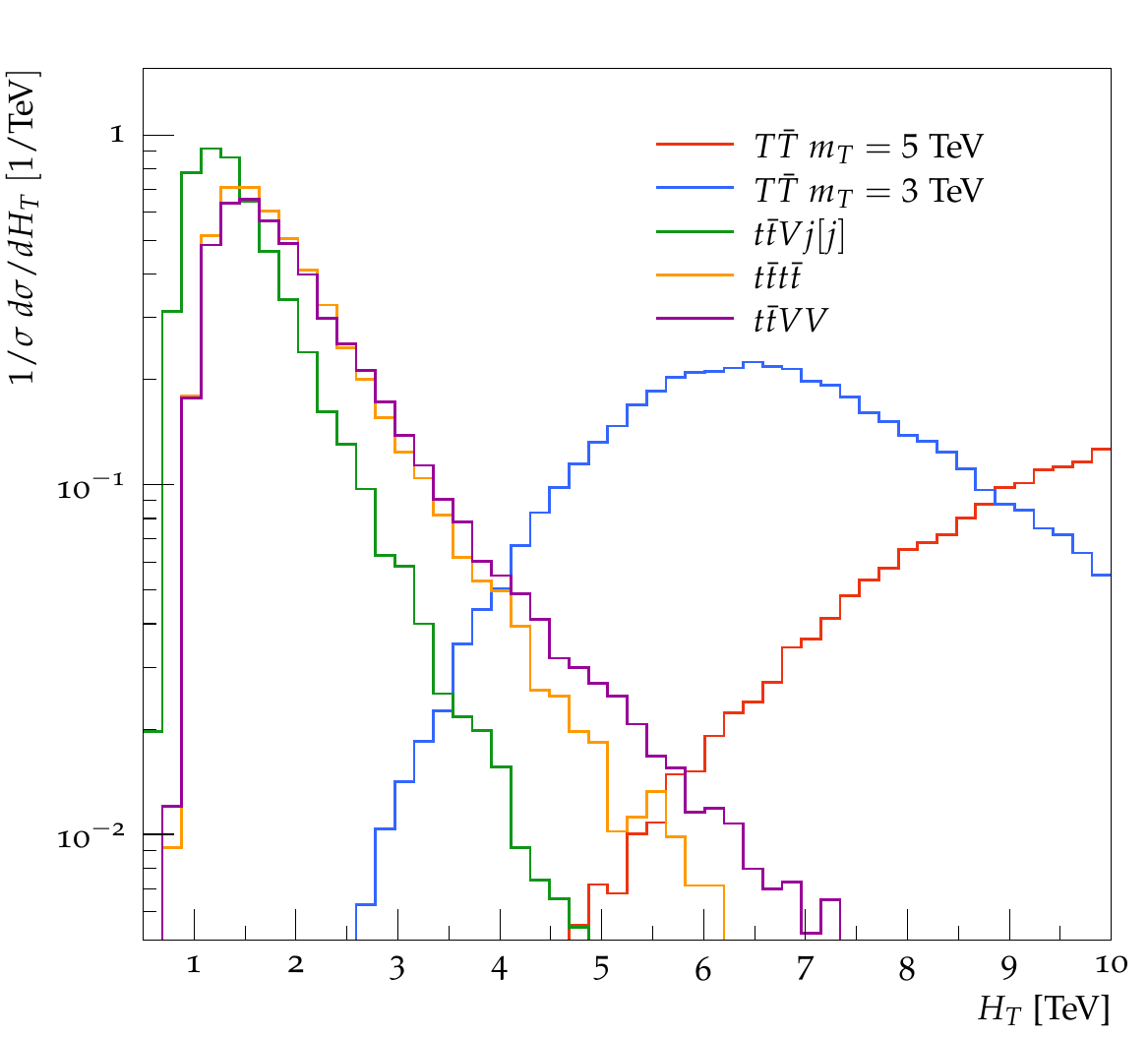}
\end{center}
\caption{$H_T$ distribution for the signal process $pp \to T T$ with $m_T=5 \text{ TeV}$ (red) and $m_T=3 \text{ TeV}$ (blue)
and the background processes $t\bar{t}VV$ (violet), $t\bar{t}t\bar{t}$ (orange) and $t\bar{t}Vj$ (green).}\label{fig:HT}
\end{figure}

We apply the following basic selection cuts:
\begin{itemize}
\item exactly three leptons with $|\eta_{\ell}|< 2.5$ and $p_{T,\ell_1}> 250\text{ GeV}$, $p_{T,\ell_2}> 100\text{ GeV}$ and $p_{T,\ell_3}> 20\text{ GeV}$
\item at least four jets with $p_{T,j}> 40\text{ GeV}$ and $|\eta_{j}|<5$
\item a cut on the transverse momentum of the leading jet $p_{T,j_1}> 70\text{ GeV}$
\item an angular separation between the jets and the leptons of \begin{equation*}\Delta R(j,\ell)= \sqrt{\Delta \phi_{j \ell}^2 +\Delta \eta_{j \ell}^2}>0.3\end{equation*}
\item
We consider leptons to be isolated if they satisfy the so-called ``mini-isolation'' criterion  \cite{Rehermann:2010vq}, as for very boosted objects the angle between the leptons and other decay products of the boosted object decreases. Hence we require for our analysis that
\begin{equation}
p_{T}^{\text{in cone}}/p_{T\ell}> 0.1 \, ,
\end{equation}
if
\begin{equation}
 \Delta R(\ell, \text{track}) < 10 \text{ GeV}/ p_{T,\ell} \, ,
 \end{equation}
and
\begin{equation}
p_{T,}^{\text{in cone}}=\sum_{\text{track}} p_{T,\text{track}}\, ,
\end{equation}
where the sum runs over all tracks (except the respective lepton track) with $p_{T,\text{track}}> 1\text{ GeV}$.
\end{itemize}
Furthermore, we apply the following cuts in the order
\begin{enumerate}
\item we require that exactly two of the jets are tagged as $b$\,--jets, $n_b=2$. The angular separation between $b$\,--jets and light jets is required to be $\Delta R(j,b)>0.3$. The tagging efficiency of the $b$\,--jets is set to 70\% and the mistagging efficiency to 2\%.
\item we set a cut on $H_T> 6 \text{ TeV}$ where $H_T=\sum_{\text{leptons}} p_{T,\ell} + \sum_{\text{jets}} p_{T,j}+ E_{T,miss} $. 
\item if we discuss the searches for 2/3 charged fermions we also request that the one $Z$ boson mass is reconstructed from either a $e^+ e^-$ pair or a $\mu^+\mu^-$ pair in the window $71 \text{ GeV} < M_{\ell^+\ell^-}< 111\text{ GeV}$.
\end{enumerate}
The cutflow for the background processes and the signal for $m_T= 5\text{ TeV}$ under the assumption that $\text{Br}(T\to tZ)=1$ is shown in Table~\ref{tab:background}. We give also the cutflow for $X_{5/3}\bar{X}_{5/3}$ pair production for $m_{X_{5/3}}= 5\text{ TeV}$ under the assumption that $\text{Br}(X_{5/3}\to W^+t)=1$. Note that this is equivalent to $B\bar{B}$ production. As it can be inferred from the table, the cut on $H_T$ gives a very good handle on the signal over the background. We exemplify this also further in Fig.~\ref{fig:HT} where the $H_T$ distribution for the different background processes and the $T\bar{T}$ signal  for the masses $m_T= 5 \text{ TeV}$ (red) and $m_T= 3\text{ TeV}$ (blue) is shown. For simplicity of the figure, we unify the background processes $t\bar{t}W^+W^-$, $t\bar{t}W^{\pm}Z$  and $t\bar{t}ZZ$ and the processes $t\bar{t}Zj[j]$ and $t\bar{t}W^{\pm}j[j]$ into one, since they have a very similar shape (with the $t\bar{t}W^{\pm}Z$ tending to slightly larger $H_T$ than $t\bar{t}W^+W^-$ and $t\bar{t}ZZ$) . As can be inferred from the figure, the $H_T$ variable can be used to distinguish very well between signal and background processes. With increasing mass of the top partner $m_T$, the $H_T$ distribution of the signal peak at higher $H_T$. 
\begin{table}
\begin{center}
\begin{tabular}{|c|c|c|c|c|} \hline
cut flow & $\sigma$ [fb]  & $n_b=2$ &  $H_{T}>6 \text{ TeV}$ & one $Z$\\ \hline \hline
 $t\bar{t}t\bar{t}$ & 3.71 & 0.706 & $9.42\cdot 10^{-3}$  & $7.09\cdot 10^{-4}$ \\ \hline
$t\bar{t}ZZ$  & 0.306 & $2.06\cdot 10^{-2}$  & $8.58\cdot 10^{-4}$& $8.42\cdot 10^{-4}$ \\ \hline
$t\bar{t}WZ$ & 0.133 & $1.26\cdot 10^{-2}$  & $1.18\cdot 10^{-3}$  & $8.90\cdot 10^{-4}$ \\ \hline
$t\bar{t}WW$ & 1.38 & 0.111 &  $3.93\cdot 10^{-3}$   & $2.22\cdot 10^{-4}$ \\ \hline
$t\bar{t}Wj[j]$ & 2.45 & 0.111 & $1.53\cdot 10^{-3}$ & $\approx$ 0 \\ \hline
$t\bar{t}Zj[j]$ & 86.8 & 2.93   & $2.14 \cdot 10^{-2}$ & $1.84\cdot 10^{-2}$ \\ \hline
$T\bar{T}$ $m_{T}= 5\text{ TeV}$ & $7.09\cdot 10^{-2}$& $2.56\cdot 10^{-2}$ & $2.53\cdot 10^{-2}$ &$2.48\cdot 10^{-2}$ \\ \hline
$X_{5/3}\bar{X}_{5/3}$ $m_{X_{5/3}}= 5\text{ TeV}$ & $6.50\cdot 10^{-2}$ & $2.56\cdot 10^{-2}$& $2.48\cdot 10^{-2}$ & -- \\ \hline
\end{tabular}
\end{center}
\caption{Cut flow for the different background processes and the signal for $m_T=5\text{ TeV}$ with $\text{BR}(T \to t Z)=1$ and $m_{X_{5/3}}=5\text{ TeV}$ with $\text{BR}(X_{5/3} \to  W^+ t)=1$.\label{tab:background}}
\end{table}

We implement a simple counting approach and hence compute the significance by
\begin{equation}
Z=\frac{S}{\sqrt{S+B}}
\end{equation}
where $S$ is the number of signal events and $B$ is the number of background events.

We then find that at a 100 TeV collider with $\mathcal{L} =300\text{ fb}^{-1}$ ($\mathcal{L} =1000\text{ fb}^{-1}$) masses of the top partner of up to $m_T= 5 \text{ TeV}$ ($m_T= 5.7 \text{ TeV}$) can be excluded at $2\sigma$, assuming $\text{BR}(T \to t Z)=1$. The discovery reach is $m_T= 3.8 \text{ TeV}$
for $\mathcal{L} =300\text{ fb}^{-1}$ and  $m_T= 4.6\text{ TeV}$
for $\mathcal{L} =1000\text{ fb}^{-1}$. For a $5/3$ or $-1/3$ charged VLQ we find that masses up to $m_{X_{5/3}}= 4.8 \text{ TeV}$ ($m_{X_{5/3}}= 5.5 \text{ TeV}$) for $\text{BR}(X_{5/3}\to W^+ t)=1$ or $\text{BR}( B\to W^- t)=1$ can be excluded. Note that the sensitivity is a bit less stringent than for the top partners, since we could not exploit the reconstruction of the leptonically decaying $Z$ boson. However, this only results in a small effect on the significance. 

\begin{figure}[t]
\begin{center}
\includegraphics[width=0.49\columnwidth]{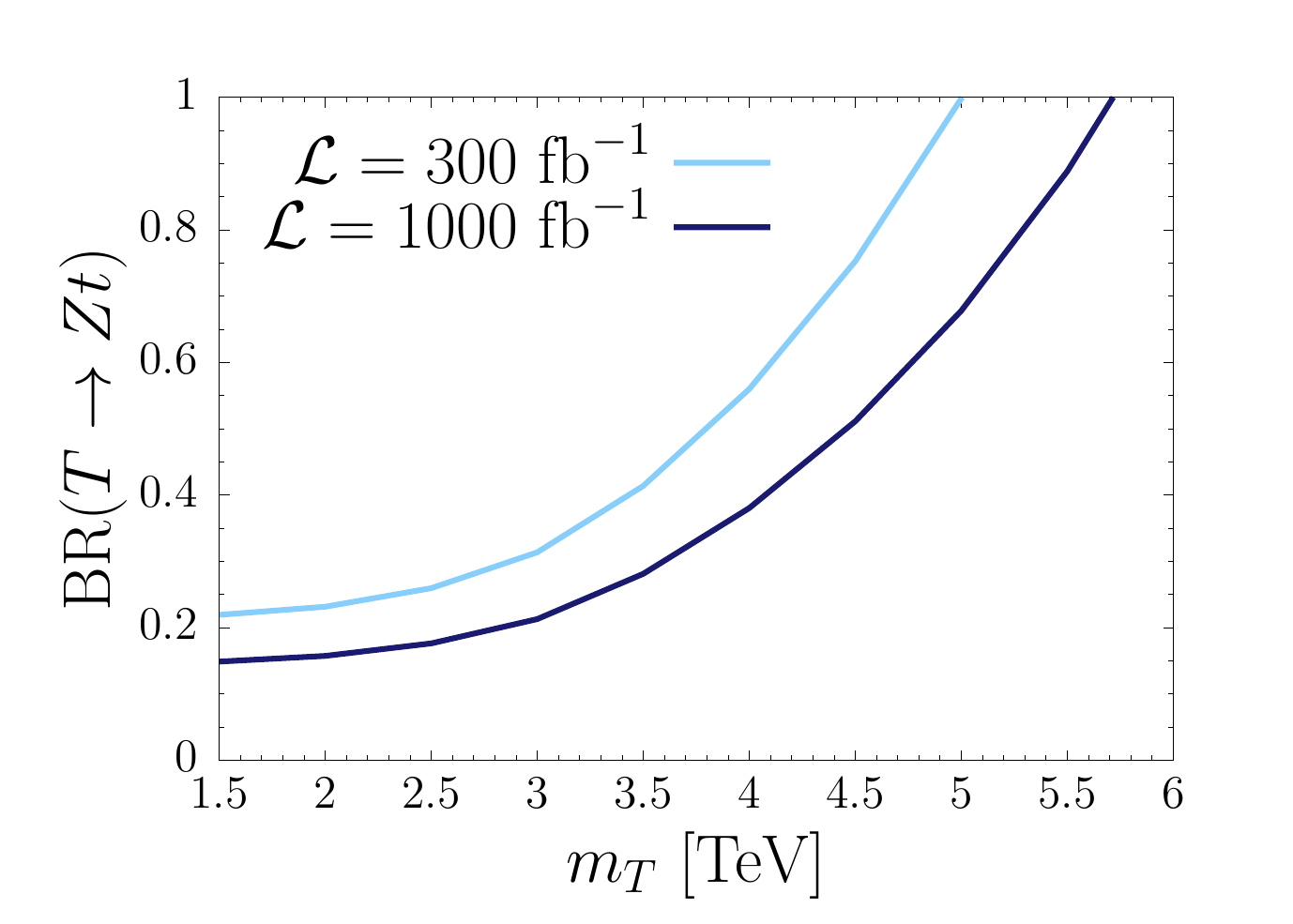}
\includegraphics[width=0.49\columnwidth]{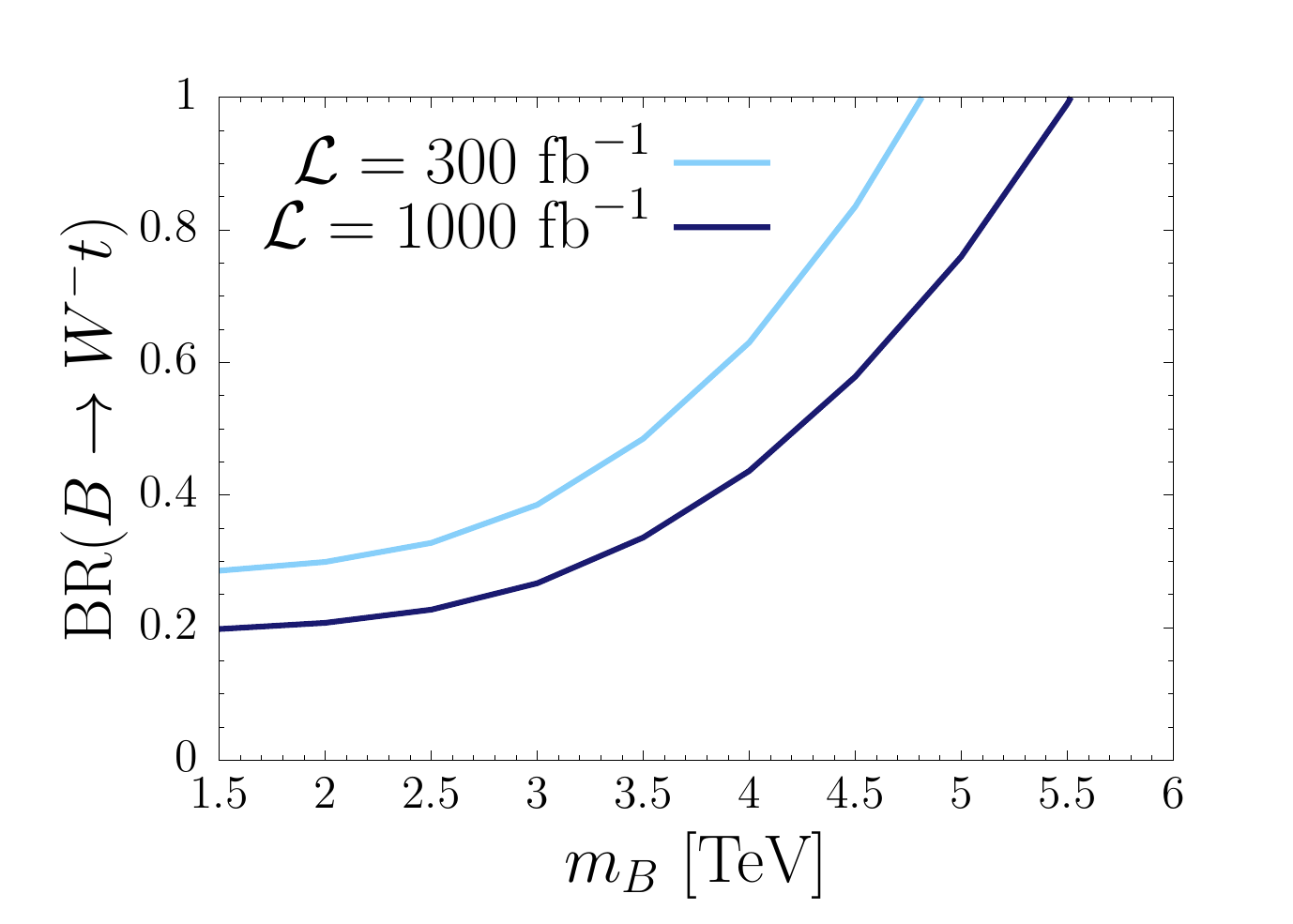}
\end{center}
\caption{BRs that can be excluded at the 2 $\sigma$ level for $T\to Zt$ (left) and $B\to W^-t$ (right) for $\mathcal{L}=300\text{ fb}^{-1}$ (light blue) and $\mathcal{L}=1000\text{ fb}^{-1}$ (dark blue) at a 100 TeV collider in the 3-lepton final state. }\label{fig:BRT}
\end{figure}

In Fig.~\ref{fig:BRT} we show the BRs that can be excluded at the $2 \sigma$ level as a function of the mass, for a top partner (left plot) and a bottom partner (right plot). Note that for low masses lower BRs can be potentially excluded if a smaller $H_T$ cut is applied. Our $H_T$ cut is optimised for large masses of the VLQ. For large BRs into other final states, \textit{e.g.}~$T\to W^+ b$, other searches are needed. Under the assumption that the BRs add up to one, this will allow to exclude also lower BRs, as shown in Fig.~\ref{fig:BRT}. A closer assessment is however beyond the scope of this paper. 

Finally, we show exclusion limits in the presence of several VLQ transforming in a $(\bf{2}, \bf{2})$ under $SU(2)_L\times SU(2)_R$ assuming their masses are approximately given by $m_{\rho}$ and their BRs are as given in Eq.~\eqref{eq:partners}. In such a case we can add up their cross sections. We then obtain that masses up to $m_{\rho}=5.7 \text{ TeV}$ ($m_{\rho}= 6.4\text{ TeV}$) can be probed for $\mathcal{L}=300\text{ fb}^{-1}$ ($\mathcal{L}=1000\text{ fb}^{-1}$).

\subsection{Search for $St \,St$ at 100 TeV}\label{sec:stops}

\begin{figure}[t]
\begin{center}
\hspace{-0.8cm}\includegraphics[width=0.5\columnwidth]{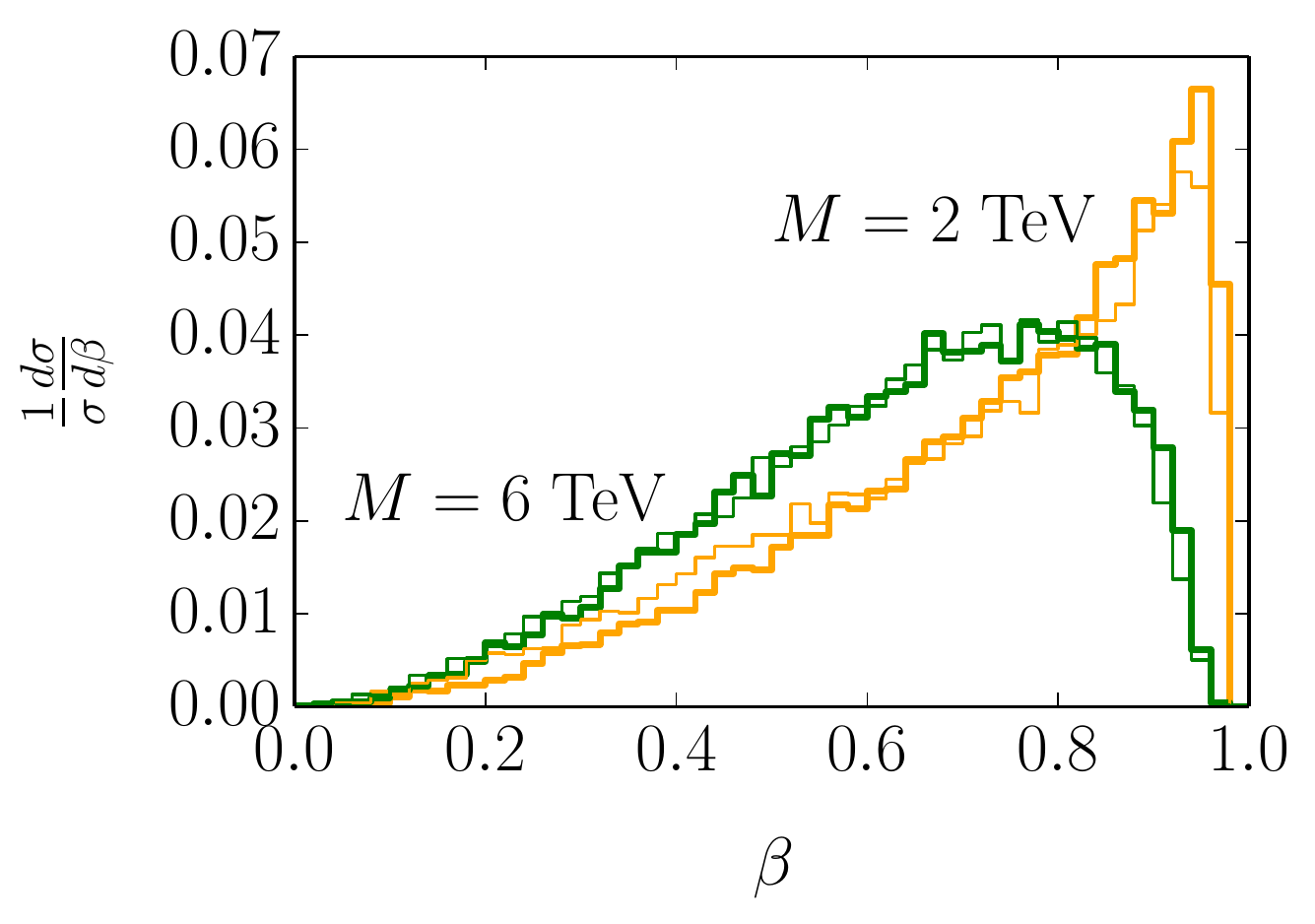}
\hspace{-0.1cm}\includegraphics[width=0.49\columnwidth]{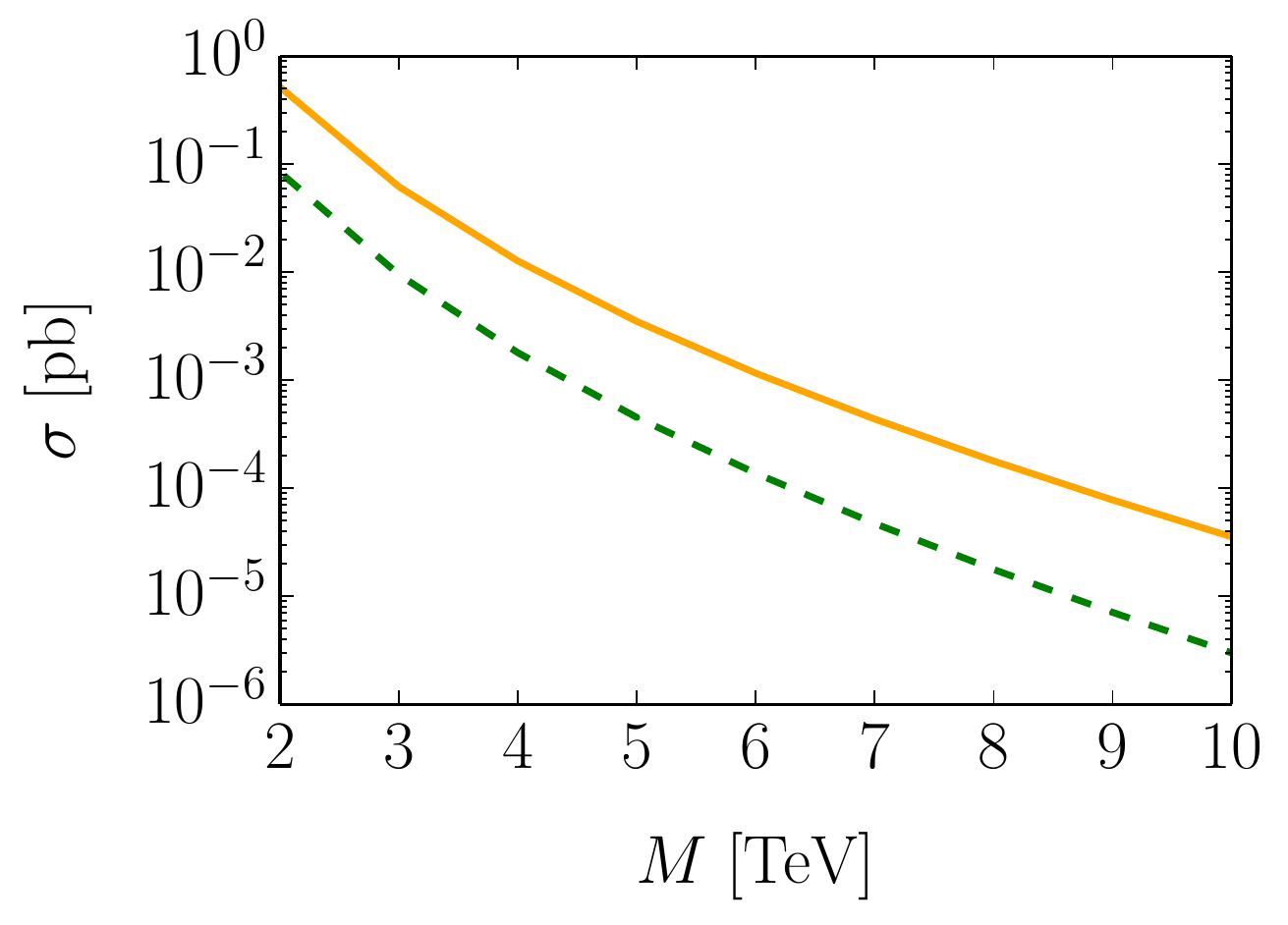}
\end{center}
\caption{\it Left) Normalized distribution for $\beta$ in pair-produced stops (thin lines) and VLQs (thick lines) for $M = 2$ TeV (orange) and $M = 6$ TeV (green). Right) Pair-production cross section at $\sqrt{s} = 100$ TeV for stops (green dashed line) and for VLQs (solid orange line). In all cases, $M$ stands for either the mass of the stop or the mass $m_\rho$ of the VLQ.}\label{fig:stopVSvlq}
\end{figure}
Prospects for $pp\to T^\prime T^\prime \to St \,St$ can be obtained from those for pair-produced stops~\cite{Cohen:2014hxa}. Although scalars and fermions present \textit{a priori} different kinematics due to the different structure of their interactions with SM particles, the kinematic differences are small. To show that, we depict in the left panel of Fig.~\ref{fig:stopVSvlq} the boost factor ($\beta$) distribution of pair-produced stops (thin line) and pair-produced VLQs (thick lines) for masses $M = 2$ TeV (orange) and $M = 6$ TeV (green). Consequently, the reach of the projected analysis in Ref.~\cite{Cohen:2014hxa} for VLQs  can be obtained by rescaling by the larger VLQ pair-production cross section. The latter is shown in the right panel of Fig.~\ref{fig:stopVSvlq} (orange solid line). The corresponding cross section for stops is also drawn (green dashed line). As it can be seen, there is almost an order of magnitude of difference between the two. 

\begin{figure}[t]
\begin{center}
\includegraphics[width=0.49\columnwidth]{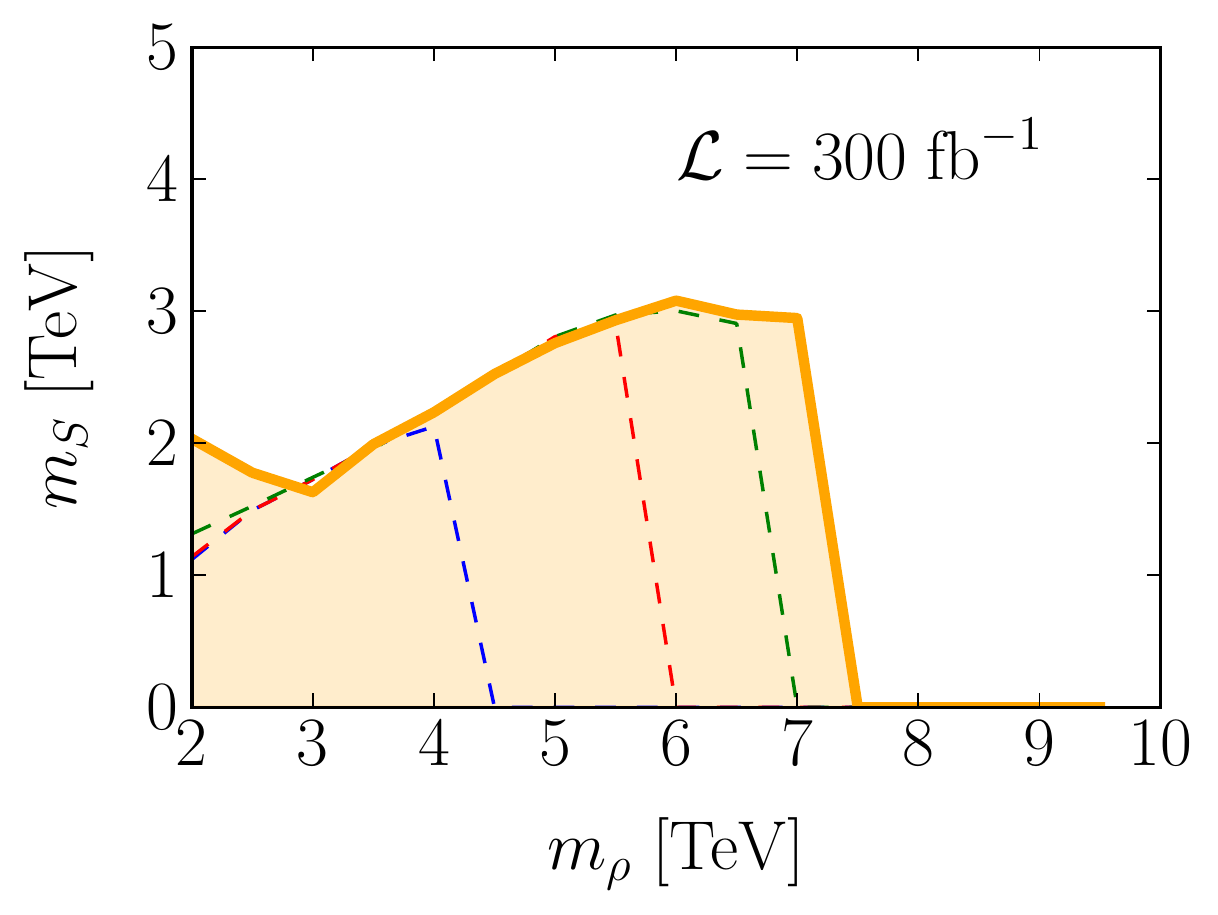}
\includegraphics[width=0.49\columnwidth]{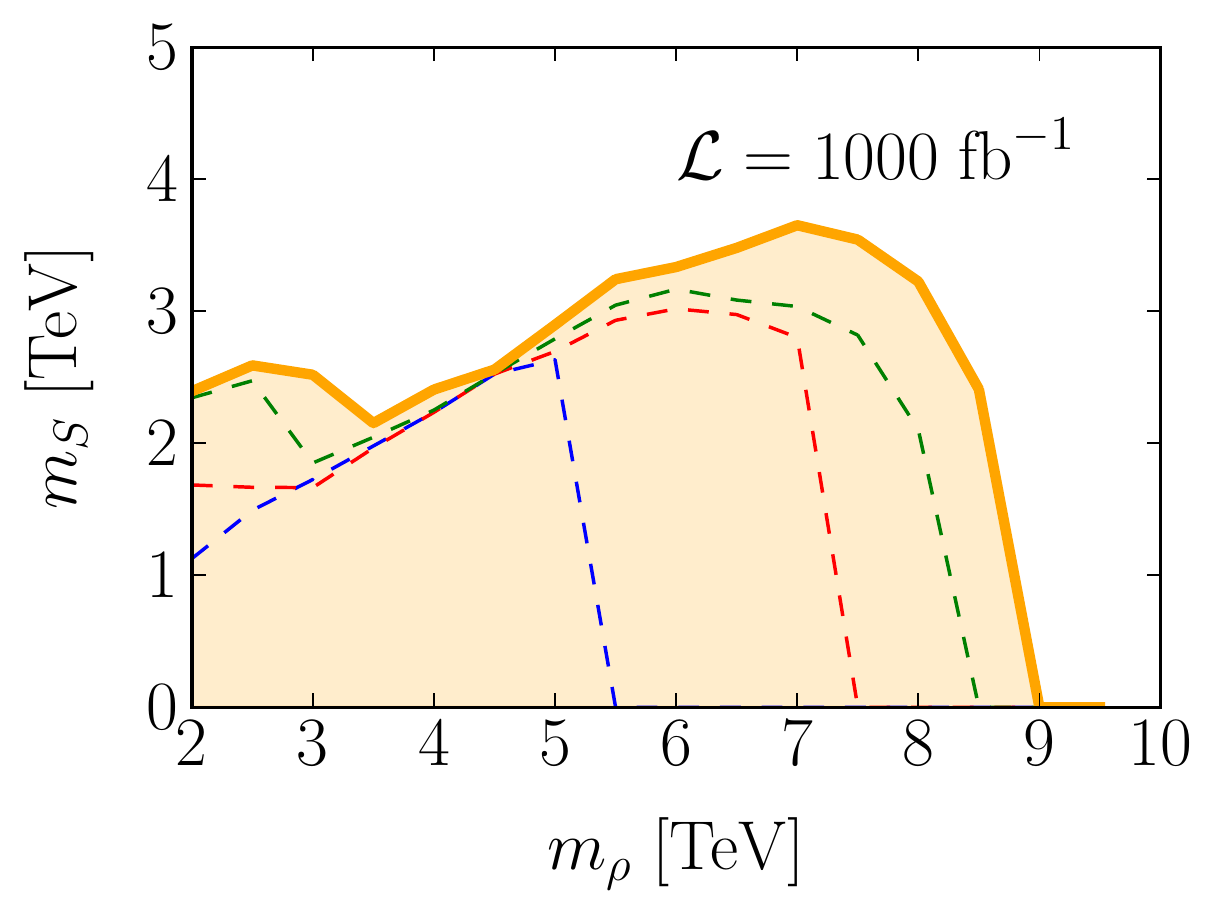}
\end{center}
\caption{\it Left) The area below the orange solid, green dashed, red dashed and blue dashed lines can be excluded at the 95 \% C.L. by recasting the analysis of Ref.~\cite{Cohen:2014hxa}  with $\mathcal{L} = 300$ fb$^{-1}$ assuming $\text{BR}(T^\prime\to St) = 1, 0.8, 0.5$ and $0.2$, respectively. Right) Same as Left) but for $\mathcal{L} = 1000$ fb$^{-1}$.}\label{fig:stst100}
\end{figure}

The concrete bound on $m_\rho$ depends also on $m_S$. We obtain the excluded regions in the plane $m_T - m_S$ in Fig.~\ref{fig:stst100}. In the left panel the integrated luminosity at a future 100 TeV collider is assumed to be $\mathcal{L} = 300$ fb$^{-1}$; in the right panel, $\mathcal{L} = 1000$ fb$^{-1}$. The regions below the solid orange line and the dashed  green, red and blue lines are excluded at the 95 \% C.L. assuming $\text{BR}(T^\prime\to St) = 1$ and $0.8, 0.5$ and $0.2$, respectively.
Note that for $\gamma$ as large as $\sim 1$, $m_S/m_T \sim 0.2$. For smaller values of $\gamma$, $m_S$ is even smaller. Thus, with $\mathcal{L} = 1000$ fb$^{-1}$, resonances of mass below $\sim 9$ TeV can be excluded.
\section{Matching to concrete models}\label{sec:matching}
In this section we consider different coset structures containing at least a Higgs doublet and an additional scalar singlet. The latter is supposed to be stabilized by an external symmetry compatible with the strong dynamics. By studying different representations in which the SM fermions (mainly the third generation $q_L$ and $t_R$) can be embedded, we will see that definite $\mathcal{O}(1)$ coefficients in Eq.~\ref{eq:parameterization} are predicted.

In order to fix the notation, let us call $T$ ($X$) the unbroken (broken) generators of any symmetry breaking pattern $\mathcal{G}\times SU(3)_c\times U(1)^\prime/\mathcal{H}\times SU(3)_c\times U(1)^\prime$. The spectator $U(1)^\prime$ is typically required to reproduce the fermion hypercharges. Hereafter we will omit both this and the colour group. Let us also define $\Pi = \Pi_a X^a$ with $\Pi_a$ running over the pNGBs. The sigma-model Lagrangian at the leading order in derivatives reads
\begin{equation}
 L_\sigma = \frac{1}{4}f^2 d^2,\quad d^2 = d_\mu^a d^{a\mu}~,
\end{equation}
where $d_\mu^a$ is the projection of the Maurer-Cartan one-form of the broken generators. It is explicitly defined by the equality
\begin{equation}
 -iU^\dagger D_\mu U = d_\mu^a X^a + T^i~\text{terms}, \quad \text{with}~ U = \exp{\left(i\sqrt{2}\frac{\Pi}{f}\right)}~.
\end{equation}

\begin{table}[hb]
\centering
\begin{tabular}{|c|c|c|c|c|c|c|}
\hline
$\mathcal{G}/\mathcal{H}$ & $q_L + t_R$ & $a_1$ & $a_2$ & $a_3$ & $\gamma$ & $\delta$  \\ \hline
%
\multirow{ 4}{*}{$SO(6)/SO(5)$} & $\mathbf{6} + \mathbf{1}$ & \multirow{ 4}{*}{$1/3$} & \multirow{ 4}{*}{$1/3$} & \multirow{ 4}{*}{$1/3$} & $-$ & $-$ \\
& $\mathbf{6} + \mathbf{15}$ &  &  &  & $\ll 1$ & $\ll 1$\\ 
& $\mathbf{15} + \mathbf{15}$ &  &  &  & $\ll 1$ & $\ll 1$\\ 
& $\mathbf{20} + \mathbf{1}$ &  &  &  & $1/4$ & $1/5$\\ \hline
\multirow{ 3}{*}{$SO(7)/SO(6)$} & $\mathbf{7} + \mathbf{1}$ & \multirow{ 3}{*}{$1/3$} & \multirow{ 3}{*}{$1/3$} & \multirow{ 3}{*}{$1/3$} & $-$ & $-$ \\ 
& $\mathbf{7} + \mathbf{7}$ &  &  &  & $-$ & $-$\\ 
& $\mathbf{27} + \mathbf{1}$ &  &  &  & $\leq 1/4$ & $\leq 1/5$\\ \hline
\multirow{ 2}{*}{$SO(7)/G_2$} & $\mathbf{8} + \mathbf{8}$ & \multirow{ 2}{*}{$1/3$} & \multirow{ 2}{*}{$1/3$} & \multirow{ 2}{*}{$1/3$} & $-$ & $-$ \\ 
& $\mathbf{35} + \mathbf{1}$ &  &  &  & $1/4$ & $1/5$\\ \hline
\multirow{ 1}{*}{$SO(6)/SO(4)$} & $\mathbf{6} + \mathbf{6}$ & \multirow{ 1}{*}{$0$} & \multirow{ 1}{*}{$1/6$} & \multirow{ 1}{*}{$1/3$} & $-$ & $-$ \\ \hline
%
\multirow{ 1}{*}{$SO(5)\times U(1)/SO(4)$} & $\mathbf{5} + \mathbf{5}$ & \multirow{ 1}{*}{$0$} & \multirow{ 1}{*}{$0$} & \multirow{ 1}{*}{$0$} & $\ll 1$ & $\ll 1$ \\  \hline
\multirow{ 1}{*}{$SO(7)/ SO(5)$} & $\mathbf{7} + \mathbf{7}$ & \multirow{ 1}{*}{$<1/3$} & \multirow{ 1}{*}{$<1/3$} & \multirow{ 1}{*}{$1/3$} & $-$ & $-$ \\  \hline
$SO(7)/ SO(6)$ &  \multirow{ 3}{*}{$\mathbf{27} + \mathbf{1}$} & \multirow{ 3}{*}{$\sim 0.3$} & \multirow{ 3}{*}{$\sim 0.3$} & \multirow{ 3}{*}{$\sim 0.3$} &  \multirow{ 3}{*}{$\sim 1/4$} &  \multirow{ 3}{*}{$\sim \sqrt{2}/5$} \\  
  &  &  &  &  &  & \\ 
$[\text{complex case}]$  &  &  &  &  &  & \\ \hline

\end{tabular}
\caption{\it Summary of the values that the $\mathcal{O}(1)$ coefficients in Eq.~\ref{eq:parameterization} can take in different CHMs. See the text in Secs.~(\ref{sec:related}-\ref{sec:complex}) for details on the possible assumptions made in each case.}
\label{tab:summary}
\end{table}
\subsection{$SO(6)/SO(5)$ and related models}
\label{sec:related}
We start considering the symmetry breaking pattern $SO(6)/SO(5)$~\cite{Gripaios:2009pe}. The generators can be chosen to be
\begin{align}\label{eq:base}
T^{mn}_{ij} &= -\frac{i}{\sqrt{2}} (\delta^m_i\delta^n_j-\delta^n_i\delta^m_j)~,
 &m<n\in[1,5]~,\\[-1mm]
X^{m6}_{ij} &= -\frac{i}{\sqrt{2}} (\delta^m_i\delta^6_j-\delta^6_i\delta^m_j)~,
 &m\in[1, 5]~.
\end{align}
$X^{16},..., X^{46}$ expand the coset space of the Higgs doublet. The broken generator associated to $S$ is provided by $X^{56}$. Expanding $L_\sigma$ in powers of $1/f$ we get
\begin{align}\nonumber
 L_\sigma &= |D_\mu H|^2 \left[1-\frac{S^2}{3f^2}\right] + \frac{1}{2}(\partial_\mu S)^2 \left[1-2\frac{|H|^2}{3f^2}\right] + \frac{1}{3f^2}\partial_\mu |H|^2 (S\partial_\mu S) + \cdots
\end{align}
where the ellipsis stands for terms with higher powers of $1/f$. This Lagrangian already fixes the values of $a_1=a_2=a_3=1/3$ in Eq.~\ref{eq:parameterization}. The exact values of the other $\mathcal{O}(1)$ parameters depend on the pairs of representations that embed the third generation $q_L$ and $t_R$.
Different choices can generate a scalar potential and the SM-like Yukawa Lagrangian without breaking the external symmetry stabilizing the scalar $S$. Among others, $q_L+t_R$ can transform in: $\mathbf{6}+\mathbf{1}$, $\mathbf{6}+\mathbf{6}$, $\mathbf{6}+\mathbf{15}$, $\mathbf{15}+\mathbf{15}$ or the $\mathbf{20}+\mathbf{1}$. Note that we are implicitly assuming that the third generation $q_L$ and $t_R$ mix with only one composite operator. This implies that lighter generations might be embedded in different representations. For example, if the choice $\mathbf{6}+\mathbf{6}$ is made for the top quark, $b_R$ cannot be embedded \mc{in the same representation,} because it should carry a different charge under $U(1)^\prime$ than the one associated to $q_L$ (the $U(1)^\prime$ charge matches the one of $t_R$). Under this assumption, the hierarchical Yukawa matrices can originate simply from the renormalization group evolution of non-hierarchical couplings in the UV~\cite{Panico:2015jxa}.

Among the highlighted choices of representations, the first one, $\mathbf{6}+\mathbf{1}$, is problematic because $t_R\sim\mathbf{1}$ does not break the global symmetry. It must be instead broken by $q_L~\sim\mathbf{6}$, which decomposes as $\mathbf{6} = \mathbf{1}+\mathbf{5}$ under $SO(5)$. Therefore, the Higgs mass would depend on only one free parameter to leading order\footnote{It is well known~\cite{Panico:2015jxa} that the number of free parameters (encoding the details of the strong dynamics) in the leading-order term in the one-loop induced potential (namely that containing the smallest number of symmetry breaking insertions) is one unit smaller than the number of $\mathcal{H}$ invariants that can be constructed out of the irreducible representations into which a particular representation of $\mathcal{G}$ decomposes. In the case under study, two such invariants exist: the $SO(5)$ singlet resulting from the product $\mathbf{5}\times\mathbf{5}$ and the one given by $\mathbf{1}\times\mathbf{1}$.~\label{ref:spurions}}. Its smallness, required to explain why the Higgs mass is much smaller than the scale $f\sim$ TeV, can only be explained at the expense of unexpected cancellations in the strong sector. 

In the second case, $\mathbf{6}+\mathbf{6}$, however, $t_R$ does break the global symmetry and partially cancel the contribution of $q_L$ to the Higgs mass. Nevertheless, \mc{the leading-order Higgs potential has only a minimum at $H=0$, so it does not drive EWSB}. Next-to-leading order terms must be considered, which not only need to be tuned to be of similar size to those generated at leading order, but they also increase the number of free parameters. Definite predictions for $\gamma$ and $\delta$ cannot be made. The same holds true for the next two cases, $\mathbf{6}+\mathbf{15}$ and $\mathbf{15}+\mathbf{15}$. However, the embedding of $t_R$ in the $\mathbf{15}$ can respect the shift symmetry of $S$ and hence make it light. Where this the case, $\gamma$ and $\delta$ would be predicted to be $\ll 1$. We will come to this point again in Section~\ref{sec:interplay}.

Finally, the choice $\mathbf{20}+\mathbf{1}$ can lead to definite predictions. Despite the fact that $t_R\sim\mathbf{1}$ does not break the global symmetry, $q_L\sim\mathbf{20}$, which decomposes as $\mathbf{1}+\mathbf{5}+\mathbf{14}$, generates both a Higgs mass term and a quartic coupling that depend on only two unknowns; see footnote~\ref{ref:spurions}. Expanding the one-loop induced potential in powers of $1/f$ and restricting to the renormalizable terms, we find
\begin{align}
V = c_1\left[2f^2|H|^2 -\frac{16}{3} |H|^4 -\frac{8}{3} S^2|H|^2\right] + c_2\left[-\frac{7}{2}f^2|H|^2\right. \nonumber\\
+ \left.\frac{19}{3}|H|^4 -2S^2 + \frac{23}{6}S^2|H|^2\right]~.
\end{align}
We stress that the small differences with the results obtained, for example in Eq.~4.11 of Ref.~\cite{Chala:2017sjk} (in their stability-preserving limit $\zeta\to 0$), are due to the fact that we are using a shift-symmetry preserving basis that does not resum higher-orders of $1/f$.
The two unknowns, $c_1$ and $c_2$, can then be traded for the Higgs mass and its quartic coupling:
\begin{equation}
 V = \mu |H|^2 + \lambda_H |H|^4 + \frac{1}{3}f^2\lambda_H S^2 + \frac{5}{18}\lambda_H S^2 |H|^2 + \mathcal{O}\left(\frac{v^2}{f^2}\right)~.
\end{equation}
Putting all together, the relevant Lagrangian reads
\begin{align}\nonumber
 L &= |D_\mu H|^2 \left[1-\frac{S^2}{3f^2}\right] + \frac{1}{3f^2}\partial_\mu |H|^2 (S\partial_\mu S)+ \frac{1}{2}(\partial_\mu S)^2 \left[1-2\frac{|H|^2}{3f^2}\right] \\
 &-\left[\frac{1}{3}f^2\lambda_H S^2 + \frac{5}{18}\lambda_H S^2 |H|^2\right]~,
\end{align}
where $\lambda_H $ denotes the usual Higgs quartic coupling.
For a mild value of $g_\rho \sim 3$, this corresponds to
\begin{equation}\label{eq:pred1}
 a_1 = a_2 = a_3 = \frac{1}{3}, \quad \gamma\sim \frac{1}{4}, \quad \delta \sim \frac{1}{5}~.
\end{equation}
Departures from this relation might appear if the gauge contribution to the scalar potential were sizable, too. 

Similar results to the ones discussed so far apply also to other models based on the coset $SO(n+1)/SO(n)$. They all develop a Higgs doublet and $n-4$ singlets. Thus, for example,
representations such as the $\mathbf{7}+\mathbf{1}$, $\mathbf{7}+\mathbf{7}$ or $\mathbf{21}+\mathbf{7}$ in $SO(7)/SO(6)$ require the inclusion of both the leading and the next-to-leading terms in the one-loop induced potential to achieve EWSB. The choice $\mathbf{27}+\mathbf{1}$, instead, gives rise to a Lagrangian very similar to the one in the equation above:
\begin{align}\label{eq:lag27}\nonumber
 L &= |D_\mu H|^2 \left[1-\frac{1}{3f^2}\left(S^2+S'^2\right)\right] + \frac{1}{2}\left[(\partial_\mu S)^2+(\partial_\mu S')^2\right] \left[1-2\frac{|H|^2}{3f^2}\right]\\
 &+ \frac{1}{3f^2}\partial_\mu |H|^2 (S\partial_\mu S + S'\partial_\mu S') -\left[\frac{1}{3}f^2\lambda_H (S^2+S'^2) + \frac{5}{18}\lambda_H (S^2+S'^2) |H|^2\right]~.
\end{align}
The phenomenology of the DM particle $S$ is thus identical to the case of $SO(7)/SO(6)$, provided that the extra pNGB, $S^\prime$, is heavier.

A mass splitting between the two singlets cannot come from gauging $SO(6)$ (which induces a potential only for the Higgs doublet). Instead, it has to arise due to global-symmetry breaking induced through the fermionic sector. For example, a small increase of the mass of the second pNGB singlet arises if $b_R$ is embedded in the appropriate irreducible representation of a $\mathbf{7}$ of $SO(7)$. Note that this representation reduces to
\begin{equation}
 \mathbf{7} = \mathbf{1} + \mathbf{6} = \mathbf{1} + \mathbf{1} + \mathbf{1} + \mathbf{4}~,
\end{equation}
under $SO(6)$ and $SO(4)$, respectively. If $b_R$ has components in the first singlet (which is a total singlet of $SO(6)$) and the second one (which is also a singlet of $SO(5)\subset SO(6)$), then the mass of the non-DM Goldstone can be increased. In the base analogous to that of Eq.~\ref{eq:base}, this embedding can be achieved by
\begin{equation}
 B_R = (0, 0, 0, 0, 0, i\zeta b_R, b_R), \quad\zeta > 0~.
\end{equation}
This gives a mass splitting of order
\begin{equation}
 m_{S^\prime}-m_{S} = \frac{m_{S^\prime}^2-m_{S}^2}{m_{S^\prime}+m_{S}} \sim \frac{m_{S^\prime}^2-m_{S}^2}{2m_{S_1}}\sim \frac{m_\rho y_b\zeta^2}{8\pi}~.
\end{equation}
Note that small differences in the coefficients of the potential would be expected if $S$ and $S'$ mix and the physical DM were a linear combination of both $S$ and $S'$. The parameters $a_1, a_2$ and $a_3$ in the derivative interactions, instead, would remain the same, given that the derivative Lagrangian is invariant under an arbitrary $SO(2)$ transformation of $S$ and $S'$.
Finally, the coset $SO(7)/G_2$ provides another very similar scenario. The pNGB spectrum consists of the Higgs doublet, a neutral scalar $S$ and singly-charged singlet $\kappa^\pm$. Among the smallest representations that can embed the third generation $q_L$ and $t_R$ we find $\mathbf{8}+\mathbf{8}$ and $\mathbf{35}+\mathbf{1}$. The second leads again to the predictions of Eq.~\ref{eq:pred1}. The DM phenomenology is, therefore, similar to that of $SO(6)/SO(5)$ with $\mathbf{20}+\mathbf{1}$, provided, again, that $\kappa^\pm$ is heavier than $S$ and hence cannot be produced in DM annihilations. Likewise in the case of $SO(7)/SO(6)$, this splitting can be triggered by $b_R$ if it is embedded, for example, in the $\mathbf{7}$ within the $\mathbf{21} = \mathbf{7}+\mathbf{14}$ of $SO(7)$. A larger splitting arises however from the hypercharge interactions (note that $\kappa^\pm$ is charged while $S$ is not). It can be estimated to be
\begin{equation}
 m_{\kappa^\pm}-m_S \sim \frac{m_\rho g'^2 }{4\pi}~.
\end{equation}
\subsection{$SO(6)/SO(4)$}\label{sec:so6so4}
This coset is special in the sense that the pNGBs transform in a \textit{reducible} representation of the unbroken group $SO(4)$, namely $\mathbf{1}+\mathbf{4}+\mathbf{4}$. In the basis of Eq.~\ref{eq:base}, these are expanded by $X^{56}$, by $X^{15}, ..., X^{45}$ and by $X^{16}, ..., X^{46}$, respectively.
The sigma-model Lagrangian can then be written to leading order in derivatives as
\begin{equation}
 L_{\sigma} = \frac{\tilde{f}^2}{4} \left[c_{11}(d^1)^2+c_{44}(d^{4})^2 + c_{4\tilde{4}}d^4\tilde{d}^4 + c_{\tilde{4}\tilde{4}}(\tilde{d}^{4})^2\right]~.
\end{equation}
Contrary to the previous cases, it depends on several free parameters, $c_{11}, c_{44}, c_{4\tilde{4}}, c_{\tilde{4}\tilde{4}}$. The different $d$ symbols correspond to
\begin{equation}
 d^1_\mu = d_\mu^{56} X^{56}, \quad d^4_\mu = d_\mu^{15} X^{15} + ... + d_\mu^{45} X^{45}, \quad \tilde{d}^4_\mu = d_\mu^{16} X^{16} + ... + d_\mu^{46} X^{46}~.
\end{equation}
They do not mix under a generic $SO(4)$ transformation.

The coefficients $a_1, a_2$ and $a_3$ in the derivative interactions in Eq.~\ref{eq:parameterization} can then get different values. As a particular scenario, let us discuss the case $c_{11}> c_{4\tilde{4}}\sim c_{\tilde{4}\tilde{4}}\gg c_{44}$. This can be interpreted as two step symmetry breaking: $SO(6)\to SO(5)\to SO(4)$. The first takes place at a scale $\sim c_{\tilde{4}\tilde{4}}f^2$ at which the second doublet can get a mass of similar order. A phenomenological study of a scenario similar to this interpretation has been given in Ref.~\cite{Sanz:2015sua}. Provided the singlet remains light, which can occur if its associated shift symmetry \mc{is only slightly broken}, the phenomenology at the scale $f = \tilde{f}\sim c_{44}$ is that described by the parametrization of Eq.~\ref{eq:parameterization}. The relevant sigma Lagrangian reads
\begin{equation}
 L_\sigma = |D_\mu H|^2
  + \frac{1}{6f^2}\partial_\mu |H|^2 (S\partial_\mu S) + \frac{1}{2}(\partial_\mu S)^2  \left[1-2\frac{|H|^2}{3f^2}\right]  \; .
\end{equation}
With $\gamma$ and $\delta$ depending on the fermion representation that we did not specify explicitly for this case, we can conclude that
\begin{equation}
 a_1 = 0, \quad a_2 = \frac{1}{6}, \quad a_3 = \frac{1}{3}~.
\end{equation}
\subsection{$SO(5)\times U(1)/SO(4)$}
This coset has been previously considered in Ref.~\cite{Gripaios:2016mmi}. The product structure of the global group makes the pNGBs transform also in a reducible representation: $\mathbf{4}+\mathbf{1}$. As a consequence, the sigma model Lagrangian is written as
\begin{equation}
 L_\sigma = \frac{1}{4}f^2 d^2 + \frac{1}{4}f_S^2\left|\partial_\mu \exp{i\frac{\sqrt{2}}{f_S}}\right|^2 = |D_\mu H|^2 + \frac{1}{2}(\partial_\mu S)^2 + \cdots
\end{equation}
where the ellipsis stand for terms with higher powers of $1/f$ and $1/f_S$ which, however, do not make $H$ and $S$ interact. Note also that in the equation above, $d$ is constructed out of the four broken generators of $SO(5)/SO(4)$. This model provides then, a DM phenomenology very similar to that of the elementary Higgs portal. It has been also pointed out that, within this scenario, the scalar potential for $S$ vanishes unless the top quark mixes with several composite operators transforming in different representations of the global group. As we have argued before, renormalization group evolution of anarchical couplings in the UV can not explain the absence of flavour-violating effects by itself. Although this problem might be circumvent by advocating extra symmetries (see for example Ref.~\cite{Chala:2017sjk}), one might still expect the top quark to mix mostly with one composite resonance. In that case, we could obtain
\begin{equation}
 a_1 = a_2 = a_3 = 0, \quad \gamma\sim\delta \ll 1~.
\end{equation}
\subsection{$SO(7)/SO(5)$}
In the same vein as $SO(6)/SO(4)$, this model develops three multiplets of the unbroken group, transforming in a reducible representation: $\mathbf{1} + \mathbf{5} + \mathbf{5}$. Following the discussion (and notation) of Section~\ref{sec:so6so4}, again in the limit $c_{11}> c_{\tilde{5}\tilde{5}} \sim c_{5\tilde{5}}\gg c_{55}$, one might expect the sigma-model Lagrangian to read
\begin{align}\nonumber
 L_\sigma  &=  \frac{1}{4}f^2 \left( c_{11} (d^1)^2+ c_{55} (d^5)^2 \right) =  |D_\mu H|^2 \left[1-\frac{1}{3f^2}S^2\right] \\
& + \frac{1}{2}\left[(\partial_\mu S')^2 + (\partial_\mu S)^2\right]\left( 1- 2\frac{|H|^2}{3f^2}\right)+ \frac{1}{6f^2}\partial_\mu |H|^2 \left(2 S\partial_\mu S+S'\partial_\mu S'\right) + ...~,
\end{align}
where $S'$ stands for the complete singlet of $SO(5)$. As in the case of $SO(7)/SO(6)$, $S$ and $S'$ could mix, for example, if they were both protected by a $\mathbb{Z}_2$ symmetry. The difference is that, in this case, the sigma-model is not invariant under a general $SO(2)$ transformation rotating $S$ into $S'$.

Assuming that the physical DM particle is mostly $S$, one obtains
\begin{equation}
 a_1 < a_2 < a_3 = \frac{1}{3}~.
\end{equation}
\subsection{Complex dark matter in $SO(7)/SO(6)$}
\label{sec:complex}
As we have mentioned before, the sigma-model Lagrangian in $SO(7)/SO(6)$ respects an additional $SO(2)$ symmetry under which $S'$ rotates into $S$. This symmetry is of course not broken by the gauge interactions (unless that $SO(2)$ is also gauged giving rise to an extended model with a $Z'$ boson). If it were also not broken by the mixings between the elementary and the composite fermions, $S$ and $S'$ would be degenerated in mass and would form a complex DM candidate. (Note that, in such a case, there is no need to assume that the strong sector is compatible with the symmetry stabilizing the DM.) A thorough study of this model in the $\mathbf{7}+\mathbf{7}$ representation has been carried out in Ref.~\cite{Balkin:2017aep}. 

Another possibility is considering, for example,  $q_L \sim \mathbf{27}, t_R \sim \mathbf{1}$. $q_L$ is explicitly embedded in
\begin{equation}
 Q_L = \left(\begin{array}{cc}
              0_{6\times 6} & \mathbf{v}^T\\
              \mathbf{v} & 0
             \end{array}\right)~, \quad \mathbf{v} = (ib_L, b_L, it_L, -t_L, 0, 0)~.
\end{equation}
An arbitrary $SO(2)$ rotation of angle $\theta$ mixing $S$ and $S'$ can be implemented by
\begin{equation}
 R(\theta)U = \exp{[i\sqrt{2}\theta X^{56}]} U  \exp{[-i\sqrt{2}\theta X^{56}]}~,
\end{equation}
which, when acting on $Q_L$, gives $R(\theta)Q_L = Q_L$. Being $Q_L$ an eigenstate of $R(\theta)$ means that the elementary-composite mixing does not break the $SO(2)$ symmetry. The relevant Lagrangian reads in this case
\begin{align}\nonumber
 L &= |D_\mu H|^2 \left[1-\frac{2|\mathbb{S}|^2}{3f^2}\right] + (\partial_\mu \mathbb{S})^2 \left[1-2\frac{|H|^2}{3f^2}\right] + \frac{1}{3f^2}\partial_\mu |H|^2 (\mathbb{S}^\ast\overleftrightarrow{\partial_\mu} \mathbb{S})\\
 &-\left[\frac{2}{3}f^2\lambda_H |\mathbb{S}|^2 + \frac{5}{9}\lambda_H \frac{|\mathbb{S}|^2 |H|^2}{f^4}\right] + \cdots
\end{align}
For a mild value of $g_\rho \sim 3$, this can be mimicked by a real scalar scenario\footnote{Note that actually there are two solutions for  $a_1 = a_2 = a_3=31/(75\,\sqrt{2})$ and $-7/(75\,\sqrt{2})$. We remark also that the reason that we can do such a matching to the real scalar case is that both derivative and potential couplings are present. Instead for a elementary singlet it is impossible because the relic abundance fixes the coupling $\delta$ which can then not be adjusted anymore for the direct detection cross section.} with:
\begin{equation}
 a_1 = a_2 = a_3 \sim 0.3, \quad \gamma \sim \frac{1}{4}, \quad \delta \sim \frac{\sqrt{2}}{5}~.
\end{equation}

A compilation of values that can be expected for the $\mathcal{O}(1)$ coefficients in Eq.~\ref{eq:parameterization} is shown in Tab.~\ref{tab:summary}. 
%
\section{Discussion}\label{sec:interplay}
%
%
\begin{figure}[t]
\begin{center}
\includegraphics[width=0.32\columnwidth]{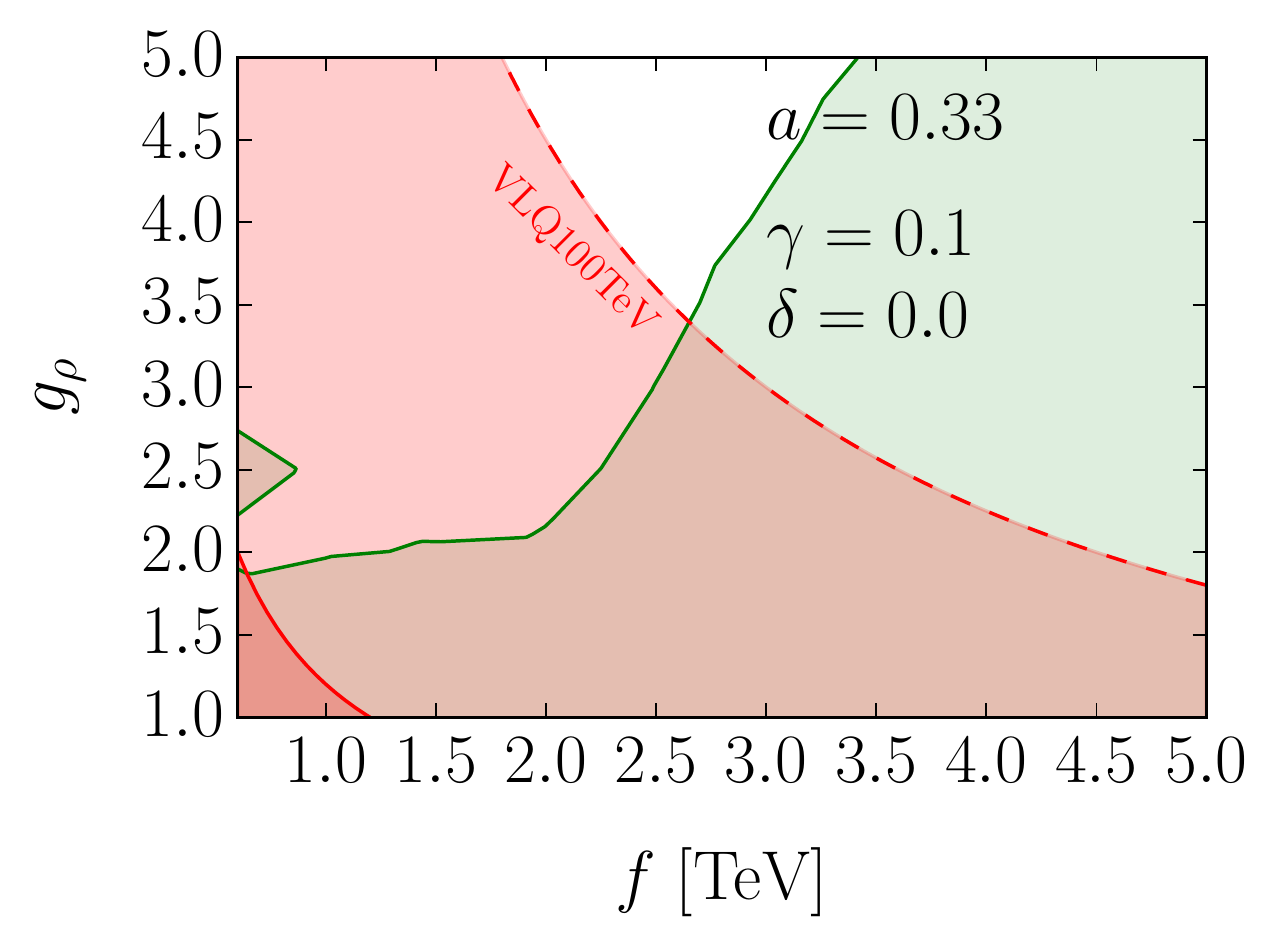}
\includegraphics[width=0.32\columnwidth]{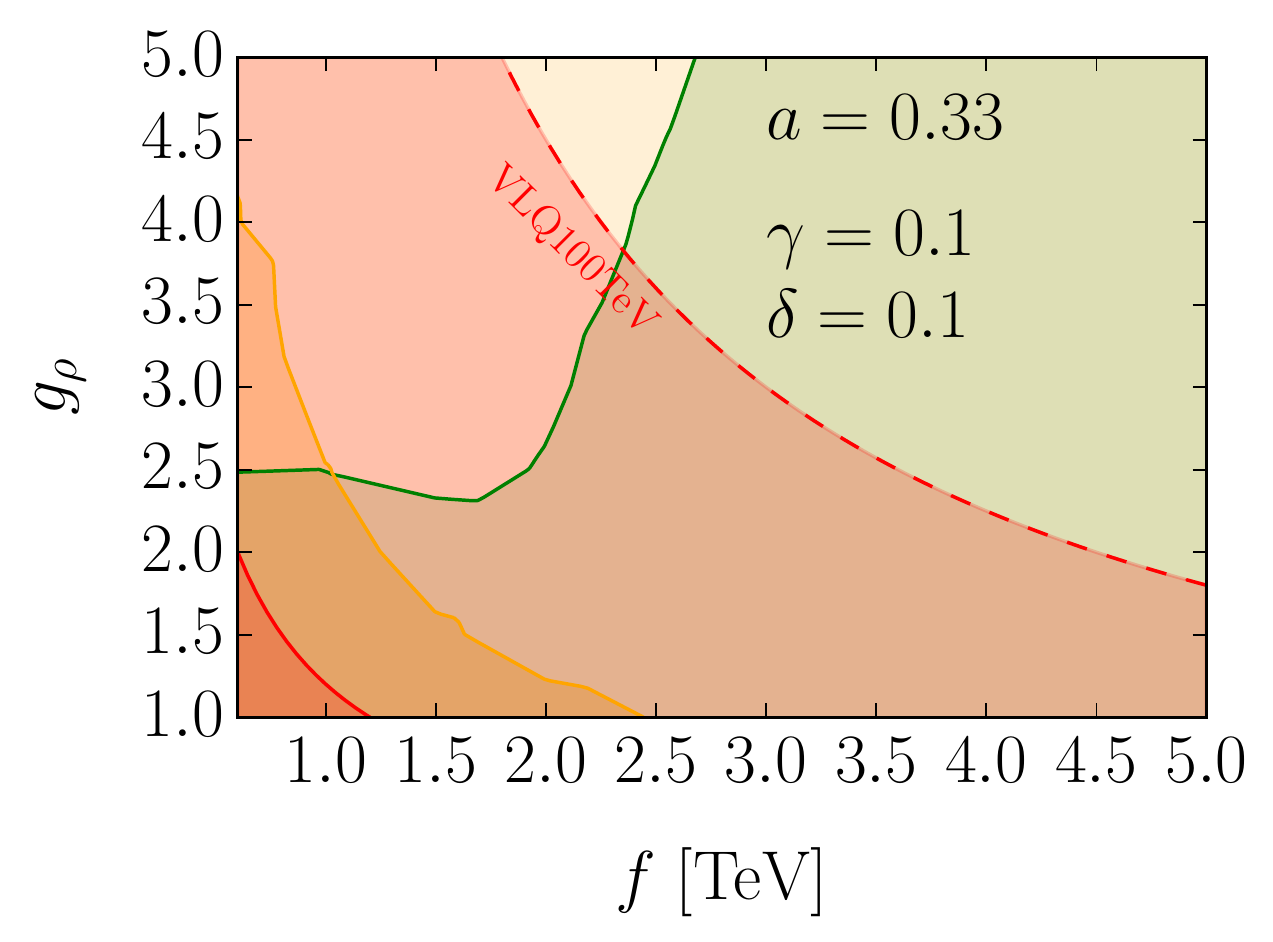}
\includegraphics[width=0.32\columnwidth]{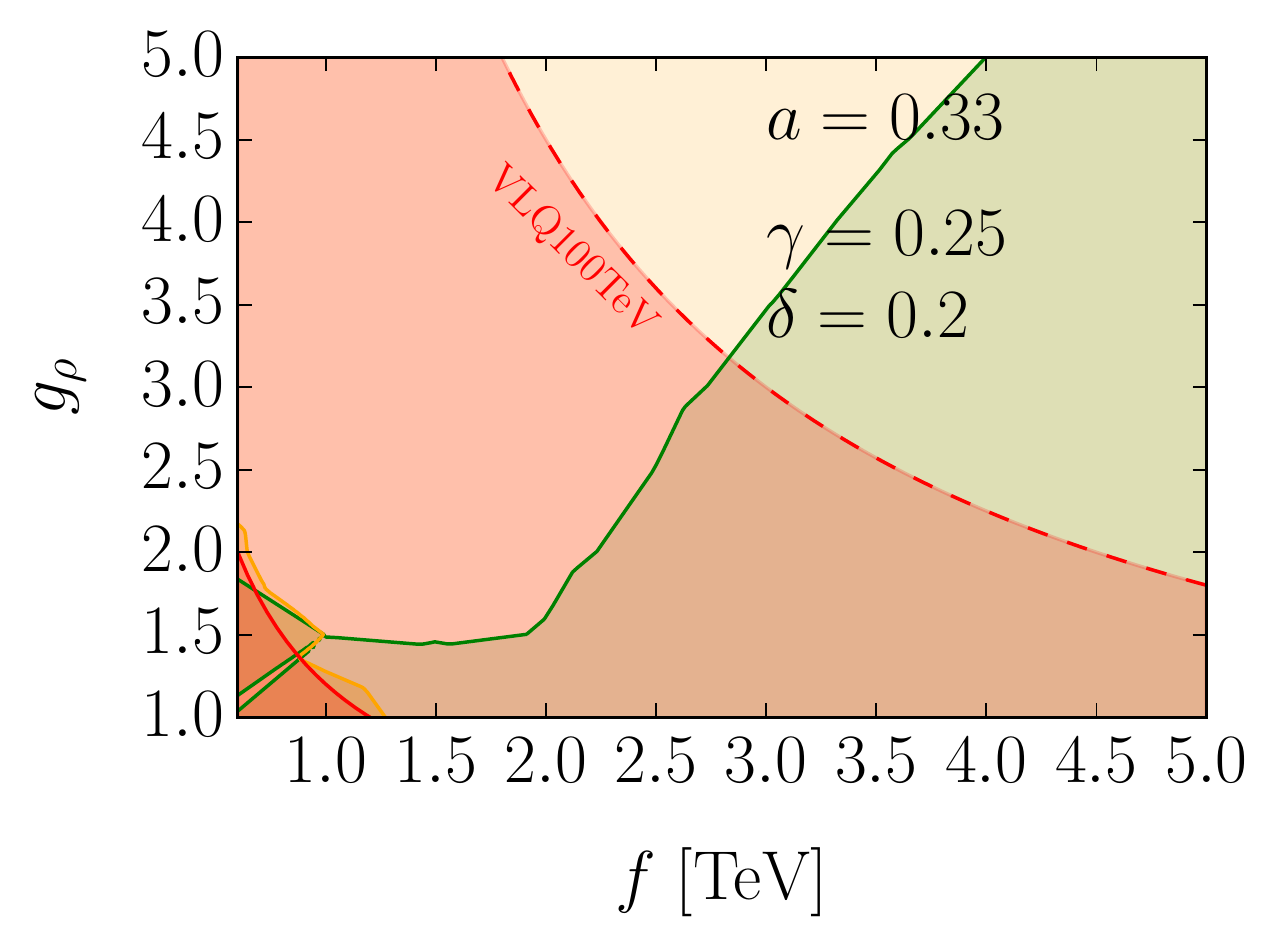}
\includegraphics[width=0.32\columnwidth]{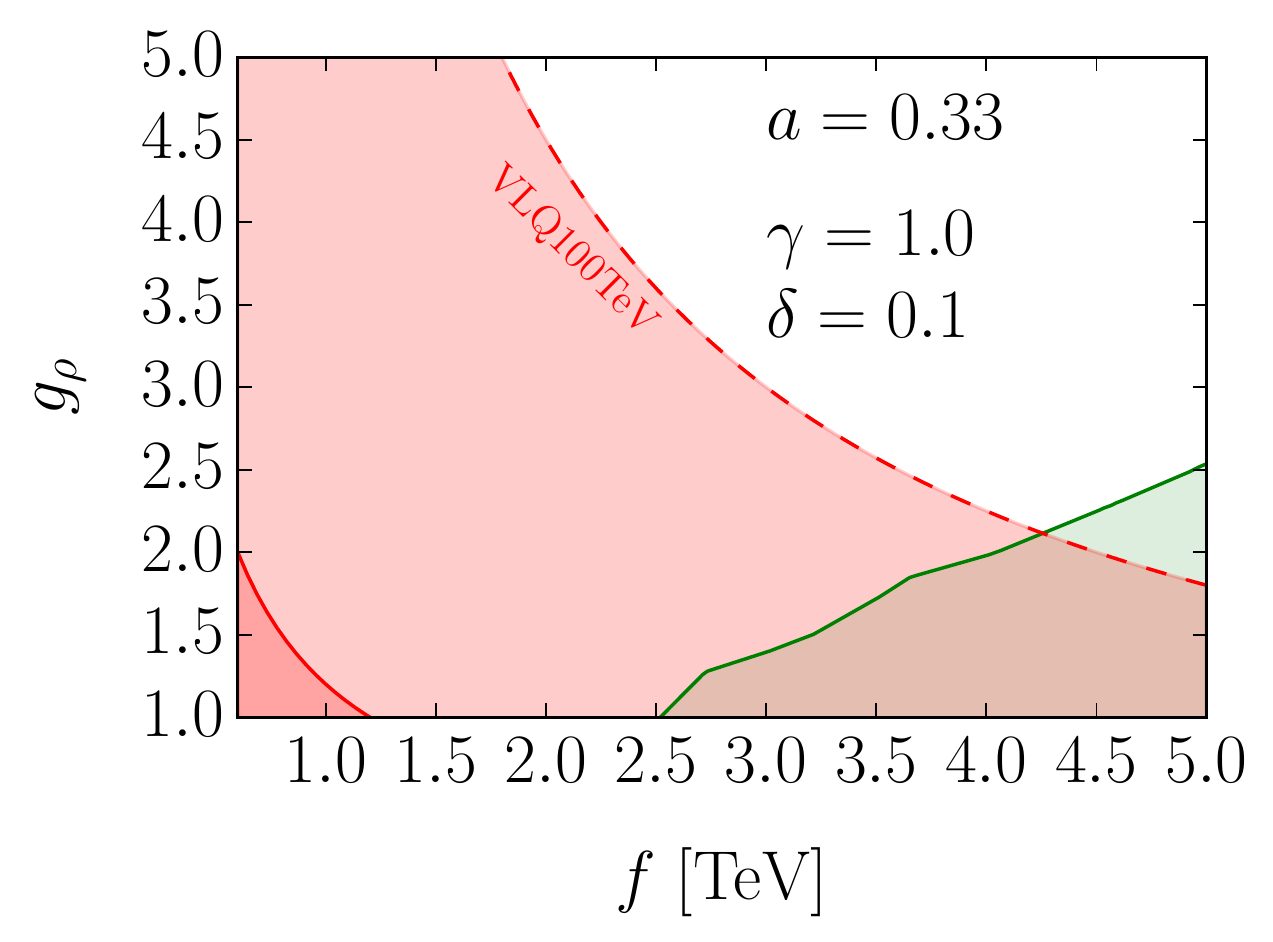}
\includegraphics[width=0.32\columnwidth]{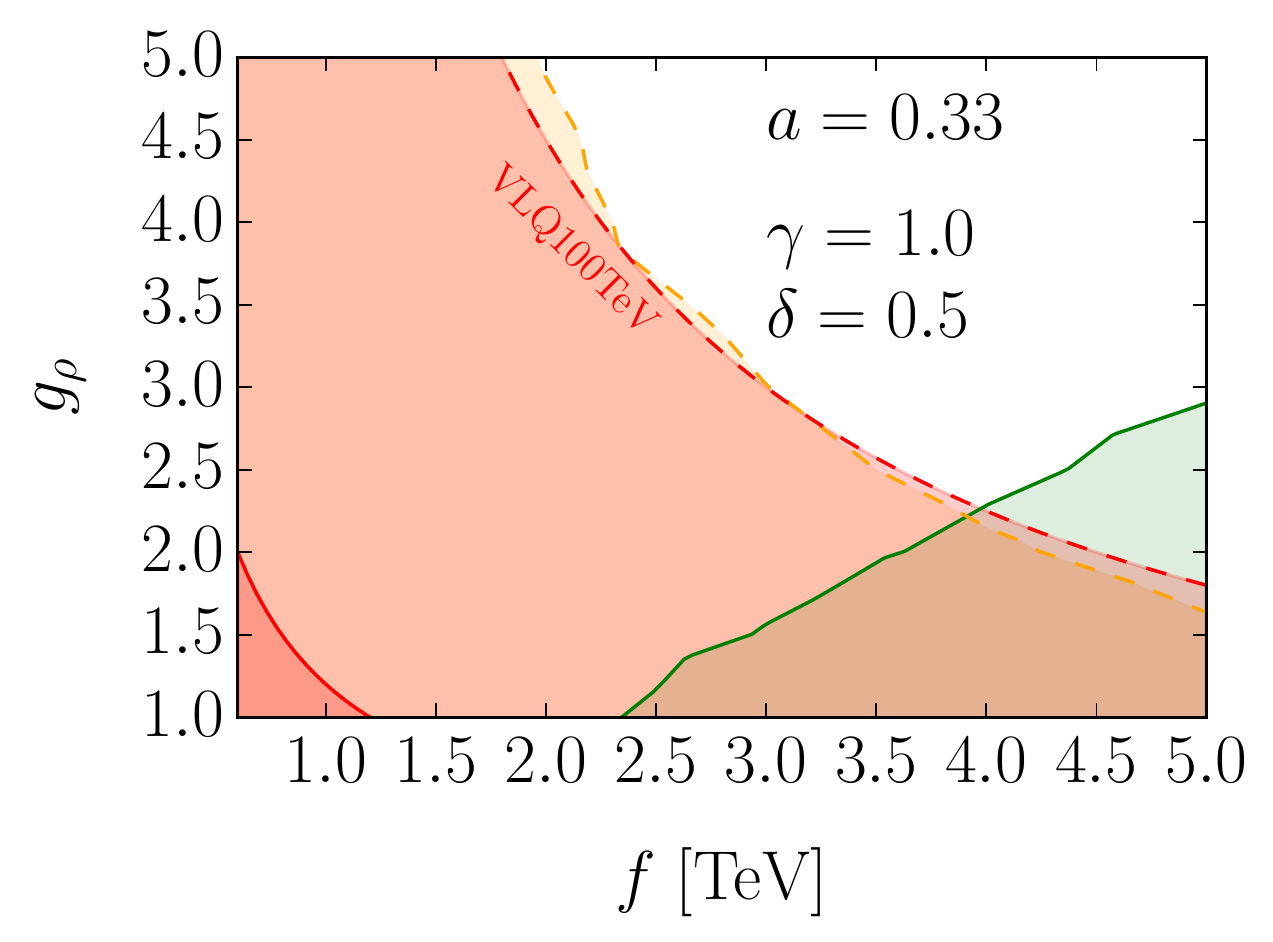}
\includegraphics[width=0.32\columnwidth]{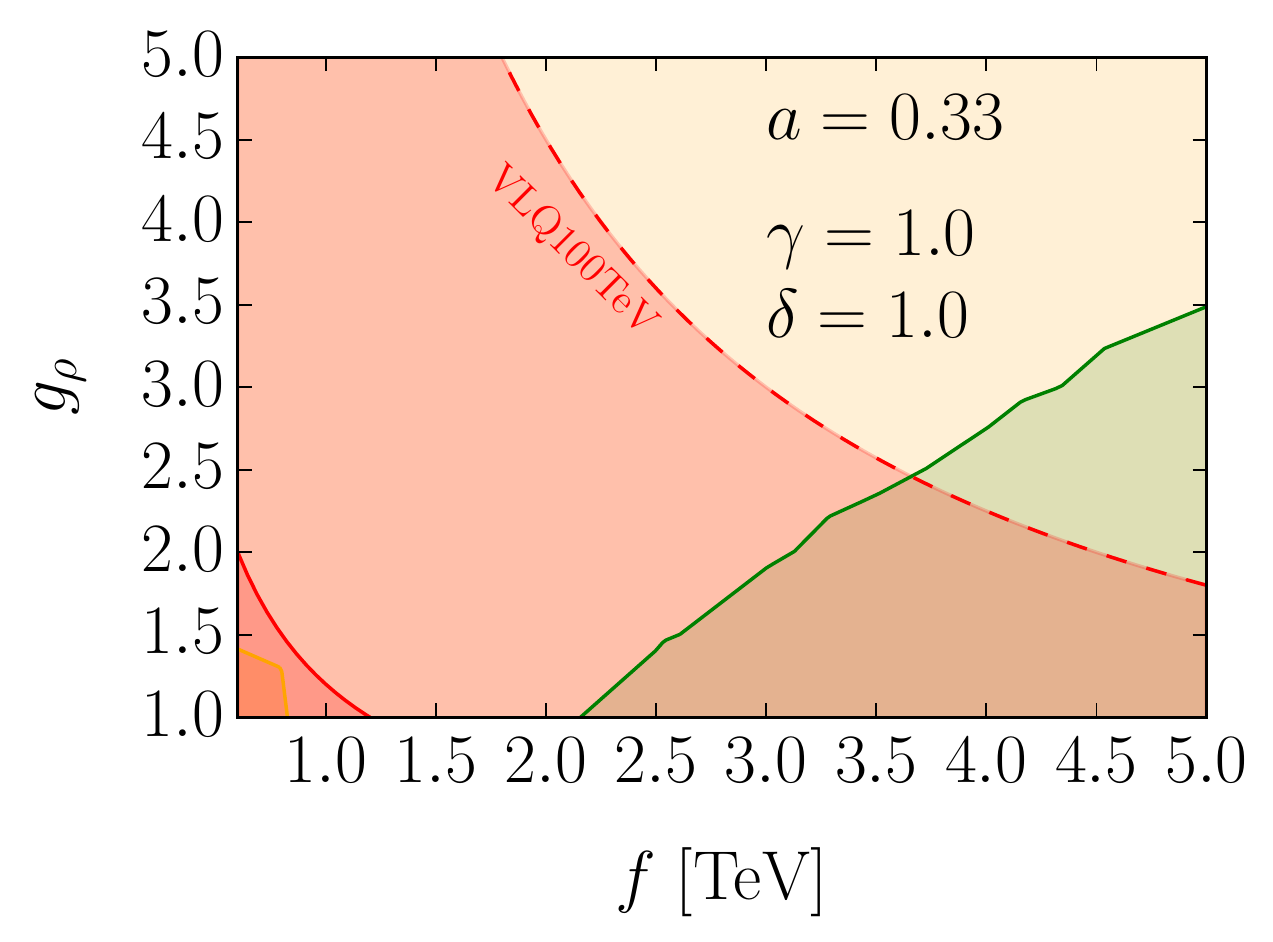}
\includegraphics[width=0.32\columnwidth]{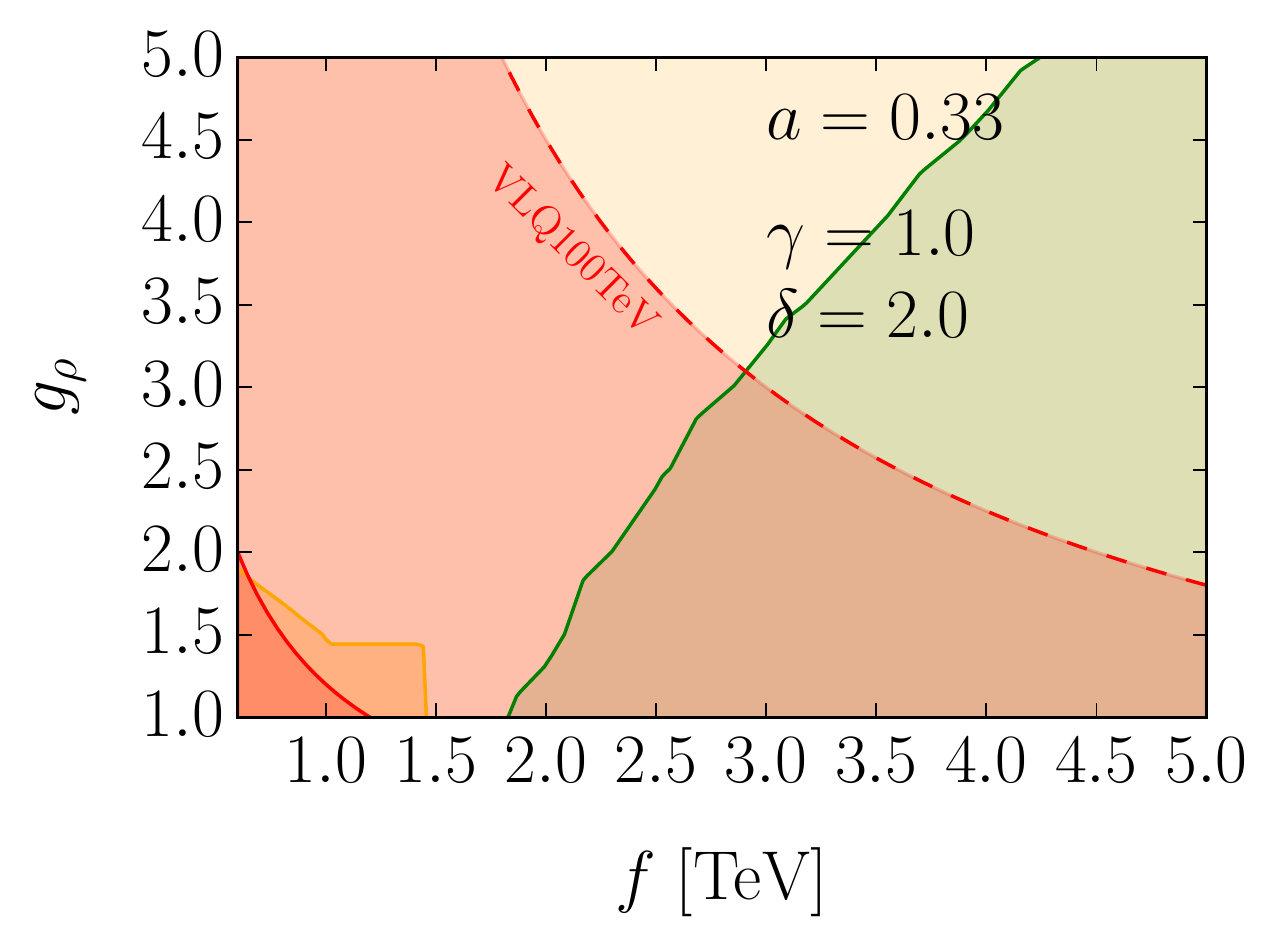}
\includegraphics[width=0.32\columnwidth]{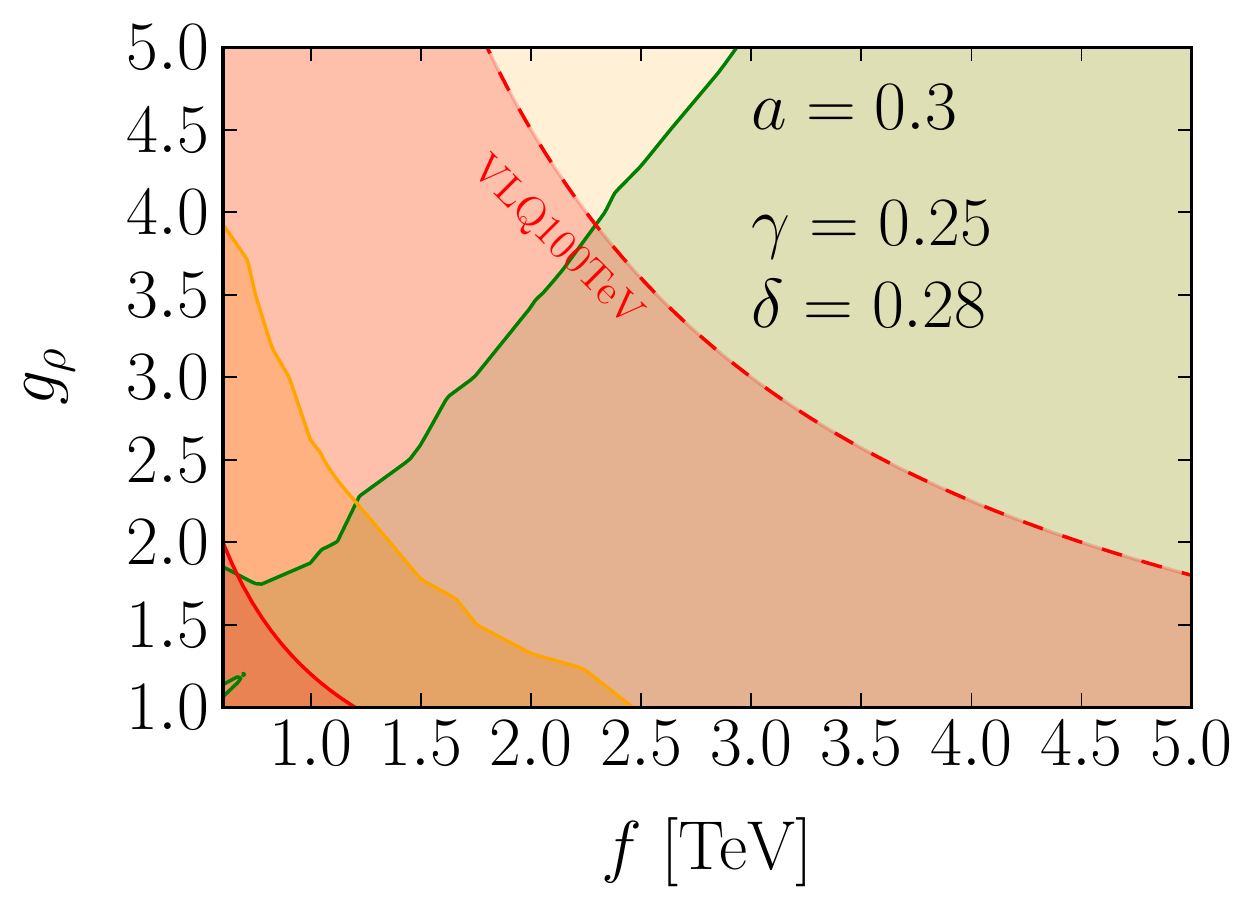}
\includegraphics[width=0.32\columnwidth]{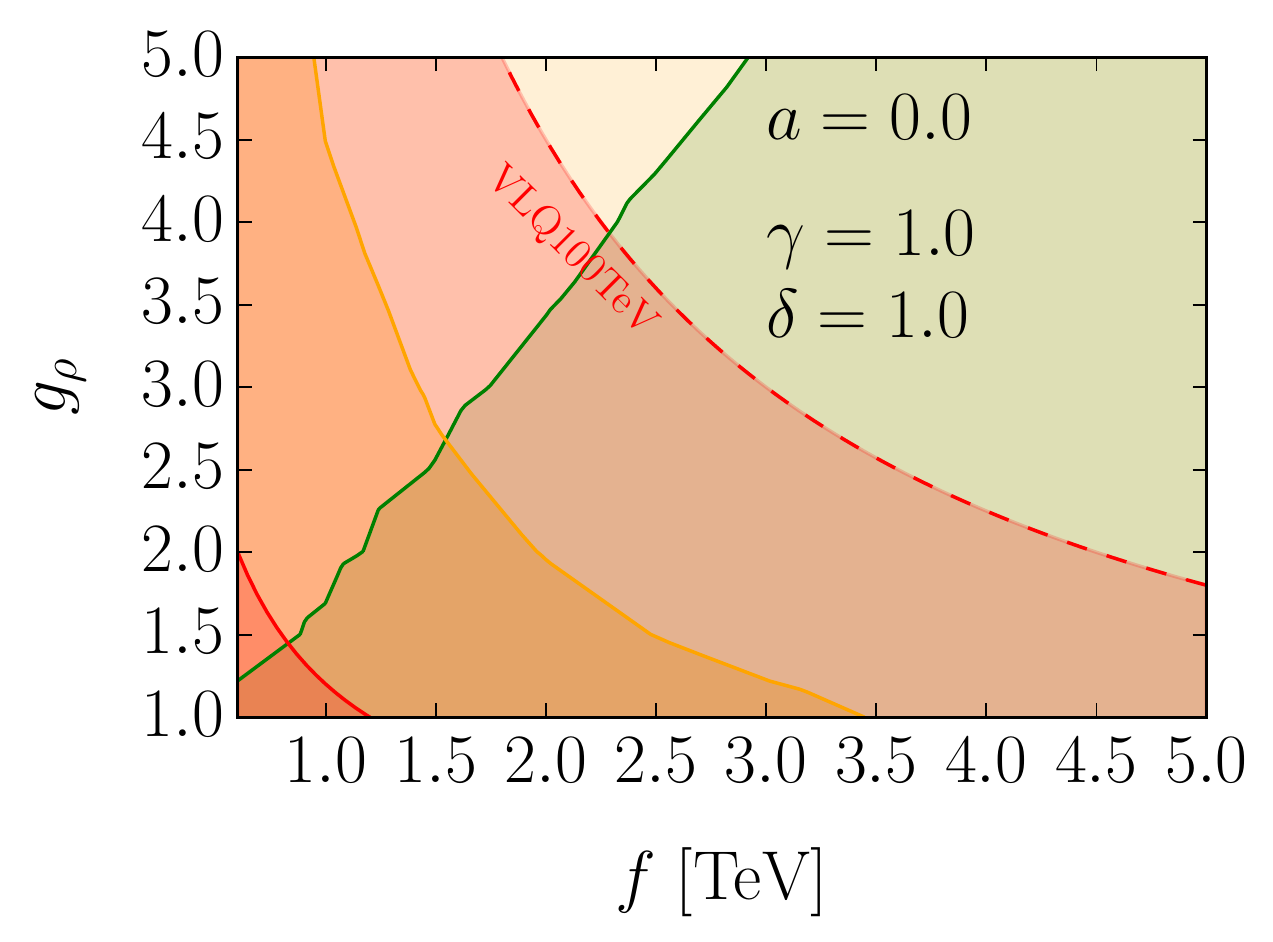}
\end{center}
\caption{\it Reach of DM and collider searches for different choices of $a\equiv (2a_1 + 2a_2 + a_3)/5$, $\gamma$ and $\delta$. In the green area enclosed by the green solid line DM is over-abundant. The orange area enclosed by the solid orange line is already excluded by LUX. The orange area enclosed by the dashed orange line will be tested by LZ. The red area enclosed by the solid red line is already excluded by the LHC. The red area enclosed by the dashed red curve can be tested by a 100 TeV collider.
}\label{fig:benchmarks}
\end{figure}
We continue by analysing the interplay between DM and collider searches for different values of the free parameters. Note that the $a_1, a_2$ and $a_3$ coefficients are mainly relevant for the computation of the relic density (which we derive using \texttt{MicrOmegas}~\cite{Belanger:2010pz}), and they enter only in the combination $2a_1 + 2a_2 + a_3$; see Eq.~\ref{eq:vertex}. We therefore define $a\equiv (2a_1 + 2a_2 + a_3)/5$. We fix this parameter, as well as $\gamma$ and $\delta$, to different benchmark points, and scan over $f$ and $g_\rho$. The results can be seen in Fig.~\ref{fig:benchmarks}. The figure can be compared to the concrete models we discussed in section~\ref{sec:matching} for the corresponding coefficients $a$, $\gamma$ and $\delta$ at $g_{\rho}=3$. Three main conclusions can be drawn:
\begin{itemize}
 \item For most values of the model-dependent coefficients, the parameter space region explaining the totality of the DM relic abundance requires $m_\rho = g_\rho f \gtrsim 2$ TeV, well above the current LHC limits. \mc{Were $\epsilon$ non-vanishing, the DM annihilation cross section would be larger, for which larger values of the DM mass would be necessary to fit the correct relic abundance. Consequently the green line in the plots above would shift to larger values of $f$, pointing to even larger $m_\rho$. We stress however again that unless there is a cancelation between the portal and derivative interactions, the operator going with $\epsilon$ is typically of small relevance.} It is also worth noting that large values of $f$ are also \textit{experimentally} excluded, due to the DM over-abundance. This gives a more robust bound than fine-tuning arguments, while still in concordance with them. 

 \item The parameter space region bounded by the measurement of the total relic density is totally complementary to that tested by direct searches for VLQs. The latter are particularly relevant when $\delta$ is small (so that direct detection experiments are poorly sensitive). Among other scenarios in this class, we find $SO(5)\times U(1)/SO(4)$, as well as models in which the $S$ shift symmetry is not broken by the top interactions, but by the bottom sector (\textit{e.g.} $SO(6)/SO(5)$ with $q_L + t_R = \mathbf{6} +\mathbf{15}$; see Tab.~\ref{tab:summary}).
 
 \item If the perturbative unitarity bound on $g_\rho$ is lowered to (the commonly used value) $\sqrt{4\pi}\sim 3.5$, almost the whole parameter space can be tested by the combination of DM and collider serches, irrespectively of the value of the $\mathcal{O}(1)$ model-dependent coefficients.
\end{itemize}
It is also worth stressing that, although not explicitly depicted in the figure, the strong bound ($m_\rho\sim 6.4$ TeV) that can be set on VLQs decaying into SM particles can be relevant if, for instance, \mc{$T'$ is much heavier than the fourplet of VLQs}. This can happen, \textit{e.g.} in $SO(6)/SO(5)$ with $q_L + t_R = \mathbf{6} + \mathbf{6}$~\cite{Serra:2015xfa}.

Finally, we note that DM scenarios in non-minimal CHMs can also be constrained by Higgs coupling measurements at future colliders. For instance in $SO(6)/SO(5)$  and related models with DM, the Higgs to vector boson couplings can be written as
\begin{equation}
g_{hVV}\approx g_{hVV}^{SM}\left(1-\frac{1}{2}\frac{v^2}{f^2}\right)~.
\end{equation}  
Using the results of Ref.~\cite{Englert:2014uua} (see also Ref.~\cite{Fujii:2015jha} for similar results) we conclude that at the ILC with energies of $250 +500 \text{ GeV}$ ($250 +500 + 1000 \text{ GeV}$) and $\mathcal{L}=250 + 500 \text{ fb}^{-1}$ ($\mathcal{L}=1150+1600+2500  \text{ fb}^{-1}$) a scale of $f=1950\text{ GeV}$ ($f=2460\text{ GeV}$) can be probed. This constraint complements the previous ones for values of $g_\rho \gtrsim 3.5$.

\section{Conclusion}
In summary, we have shown that collider searches for vector-like quarks at the LHC and future colliders can bound the parameter region that is complementary to the one bounded by the measurement of the relic density. We have provided estimates for the potential sensitivity for searches of VLQs at a 100 TeV collider. While the searches for final states with dark matter have the strongest reach, up to $m_{\rho}=9 \text{ TeV}$, for masses of the dark matter particle of up to 3 TeV, our results for the decays into SM particles only depend on the respective branching ratio and can be applied to scenarios where the lightest VLQ does not decay into the dark matter particle. In particular, we found for the 3-lepton final state that $m_{\rho}< 6.4$ TeV can be excluded for a four-plet of $SU(2)_L\times SU(2)_R$. 
In addition, we checked the limits from the direct detection of dark matter at the LUX experiment and provided estimates of the sensitivity of the future experiment LZ. In order to do so, we advocated the usage of an effective parametrization that allowed us to apply our results to a multitude of models. Taking into account all these ingredients we have shown that dark matter scenarios within Composite Higgs Models can be well probed at future experiments, leaving no or only little parameter space unexplored.

\bigskip

\section*{Acknowledgements}
We would like to thank Johannes Bellm, Frank Krauss, Jonas Lindert and Oscar Ochoa Valeriano for valuable discussions. RG is supported by a European Union COFUND/Durham Junior Research Fellowship under the EU grant number 609412. MC is supported by the Royal Society under the the Newton International Fellowships programm. MC thanks the Hong Kong Institute for Advanced Study for the hospitality during the last part of this work.


\bibliographystyle{JHEP}
\bibliography{references}

\providecommand{\href}[2]{#2}\begingroup\raggedright\begin{thebibliography}{10}

\bibitem{Dimopoulos:1981xc}
S.~Dimopoulos and J.~Preskill, \emph{{Massless Composites With Massive
  Constituents}},
  \href{http://dx.doi.org/10.1016/0550-3213(82)90345-5}{\emph{Nucl. Phys.} {\bf
  B199} (1982) 206--222}.

\bibitem{Kaplan:1983fs}
D.~B. Kaplan and H.~Georgi, \emph{{SU(2) x U(1) Breaking by Vacuum
  Misalignment}},
  \href{http://dx.doi.org/10.1016/0370-2693(84)91177-8}{\emph{Phys. Lett.} {\bf
  B136} (1984) 183--186}.

\bibitem{Kaplan:1983sm}
D.~B. Kaplan, H.~Georgi and S.~Dimopoulos, \emph{{Composite Higgs Scalars}},
  \href{http://dx.doi.org/10.1016/0370-2693(84)91178-X}{\emph{Phys. Lett.} {\bf
  B136} (1984) 187--190}.

\bibitem{Agashe:2004rs}
K.~Agashe, R.~Contino and A.~Pomarol, \emph{{The Minimal composite Higgs
  model}}, \href{http://dx.doi.org/10.1016/j.nuclphysb.2005.04.035}{\emph{Nucl.
  Phys.} {\bf B719} (2005) 165--187},
  [\href{https://arxiv.org/abs/hep-ph/0412089}{{\tt hep-ph/0412089}}].

\bibitem{Chala:2017xgc}
M.~Chala, \emph{{Direct Bounds on Heavy Top-Like Quarks With Standard and
  Exotic Decays}},  \href{https://arxiv.org/abs/1705.03013}{{\tt 1705.03013}}.

\bibitem{Araque:2016jrb}
{\scshape ATLAS} collaboration, J.~P. Araque, \emph{{Overview of the
  vector-like quark searches with the LHC data collected by the ATLAS
  detector}},  in \emph{{9th International Workshop on Top Quark Physics (TOP
  2016) Olomouc, Czech Republic, September 19-23, 2016}}, 2016.
\newblock \href{https://arxiv.org/abs/1611.09056}{{\tt 1611.09056}}.

\bibitem{Aaboud:2017zfn}
{\scshape ATLAS} collaboration, M.~Aaboud et~al., \emph{{Search for pair
  production of heavy vector-like quarks decaying to high-p$_{T}$ W bosons and
  b quarks in the lepton-plus-jets final state in pp collisions at $
  \sqrt{s}=13 $ TeV with the ATLAS detector}},
  \href{http://dx.doi.org/10.1007/JHEP10(2017)141}{\emph{JHEP} {\bf 10} (2017)
  141}, [\href{https://arxiv.org/abs/1707.03347}{{\tt 1707.03347}}].

\bibitem{Aaboud:2017qpr}
{\scshape ATLAS} collaboration, M.~Aaboud et~al., \emph{{Search for pair
  production of vector-like top quarks in events with one lepton, jets, and
  missing transverse momentum in $ \sqrt{s}=13 $ TeV $pp$ collisions with the
  ATLAS detector}},
  \href{http://dx.doi.org/10.1007/JHEP08(2017)052}{\emph{JHEP} {\bf 08} (2017)
  052}, [\href{https://arxiv.org/abs/1705.10751}{{\tt 1705.10751}}].

\bibitem{Contino:2006qr}
R.~Contino, L.~Da~Rold and A.~Pomarol, \emph{{Light custodians in natural
  composite Higgs models}},
  \href{http://dx.doi.org/10.1103/PhysRevD.75.055014}{\emph{Phys. Rev.} {\bf
  D75} (2007) 055014}, [\href{https://arxiv.org/abs/hep-ph/0612048}{{\tt
  hep-ph/0612048}}].

\bibitem{Matsedonskyi:2012ym}
O.~Matsedonskyi, G.~Panico and A.~Wulzer, \emph{{Light Top Partners for a Light
  Composite Higgs}},
  \href{http://dx.doi.org/10.1007/JHEP01(2013)164}{\emph{JHEP} {\bf 01} (2013)
  164}, [\href{https://arxiv.org/abs/1204.6333}{{\tt 1204.6333}}].

\bibitem{Redi:2012ha}
M.~Redi and A.~Tesi, \emph{{Implications of a Light Higgs in Composite
  Models}}, \href{http://dx.doi.org/10.1007/JHEP10(2012)166}{\emph{JHEP} {\bf
  10} (2012) 166}, [\href{https://arxiv.org/abs/1205.0232}{{\tt 1205.0232}}].

\bibitem{Marzocca:2012zn}
D.~Marzocca, M.~Serone and J.~Shu, \emph{{General Composite Higgs Models}},
  \href{http://dx.doi.org/10.1007/JHEP08(2012)013}{\emph{JHEP} {\bf 08} (2012)
  013}, [\href{https://arxiv.org/abs/1205.0770}{{\tt 1205.0770}}].

\bibitem{Pomarol:2012qf}
A.~Pomarol and F.~Riva, \emph{{The Composite Higgs and Light Resonance
  Connection}}, \href{http://dx.doi.org/10.1007/JHEP08(2012)135}{\emph{JHEP}
  {\bf 08} (2012) 135}, [\href{https://arxiv.org/abs/1205.6434}{{\tt
  1205.6434}}].

\bibitem{Panico:2012uw}
G.~Panico, M.~Redi, A.~Tesi and A.~Wulzer, \emph{{On the Tuning and the Mass of
  the Composite Higgs}},
  \href{http://dx.doi.org/10.1007/JHEP03(2013)051}{\emph{JHEP} {\bf 03} (2013)
  051}, [\href{https://arxiv.org/abs/1210.7114}{{\tt 1210.7114}}].

\bibitem{Azatov:2013hya}
A.~Azatov, M.~Salvarezza, M.~Son and M.~Spannowsky, \emph{{Boosting Top Partner
  Searches in Composite Higgs Models}},
  \href{http://dx.doi.org/10.1103/PhysRevD.89.075001}{\emph{Phys. Rev.} {\bf
  D89} (2014) 075001}, [\href{https://arxiv.org/abs/1308.6601}{{\tt
  1308.6601}}].

\bibitem{Ortiz:2014iza}
N.~Gutierrez~Ortiz, J.~Ferrando, D.~Kar and M.~Spannowsky,
  \emph{{Reconstructing singly produced top partners in decays to Wb}},
  \href{http://dx.doi.org/10.1103/PhysRevD.90.075009}{\emph{Phys. Rev.} {\bf
  D90} (2014) 075009}, [\href{https://arxiv.org/abs/1403.7490}{{\tt
  1403.7490}}].

\bibitem{Matsedonskyi:2015dns}
O.~Matsedonskyi, G.~Panico and A.~Wulzer, \emph{{Top Partners Searches and
  Composite Higgs Models}},
  \href{http://dx.doi.org/10.1007/JHEP04(2016)003}{\emph{JHEP} {\bf 04} (2016)
  003}, [\href{https://arxiv.org/abs/1512.04356}{{\tt 1512.04356}}].

\bibitem{Frigerio:2012uc}
M.~Frigerio, A.~Pomarol, F.~Riva and A.~Urbano, \emph{{Composite Scalar Dark
  Matter}}, \href{http://dx.doi.org/10.1007/JHEP07(2012)015}{\emph{JHEP} {\bf
  07} (2012) 015}, [\href{https://arxiv.org/abs/1204.2808}{{\tt 1204.2808}}].

\bibitem{Marzocca:2014msa}
D.~Marzocca and A.~Urbano, \emph{{Composite Dark Matter and LHC Interplay}},
  \href{http://dx.doi.org/10.1007/JHEP07(2014)107}{\emph{JHEP} {\bf 07} (2014)
  107}, [\href{https://arxiv.org/abs/1404.7419}{{\tt 1404.7419}}].

\bibitem{Fonseca:2015gva}
N.~Fonseca, R.~Zukanovich~Funchal, A.~Lessa and L.~Lopez-Honorez, \emph{{Dark
  Matter Constraints on Composite Higgs Models}},
  \href{http://dx.doi.org/10.1007/JHEP06(2015)154}{\emph{JHEP} {\bf 06} (2015)
  154}, [\href{https://arxiv.org/abs/1501.05957}{{\tt 1501.05957}}].

\bibitem{Gripaios:2009pe}
B.~Gripaios, A.~Pomarol, F.~Riva and J.~Serra, \emph{{Beyond the Minimal
  Composite Higgs Model}},
  \href{http://dx.doi.org/10.1088/1126-6708/2009/04/070}{\emph{JHEP} {\bf 04}
  (2009) 070}, [\href{https://arxiv.org/abs/0902.1483}{{\tt 0902.1483}}].

\bibitem{Chala:2016ykx}
M.~Chala, G.~Nardini and I.~Sobolev, \emph{{Unified explanation for dark matter
  and electroweak baryogenesis with direct detection and gravitational wave
  signatures}}, \href{http://dx.doi.org/10.1103/PhysRevD.94.055006}{\emph{Phys.
  Rev.} {\bf D94} (2016) 055006}, [\href{https://arxiv.org/abs/1605.08663}{{\tt
  1605.08663}}].

\bibitem{Balkin:2017aep}
R.~Balkin, M.~Ruhdorfer, E.~Salvioni and A.~Weiler, \emph{{Charged Composite
  Scalar Dark Matter}},
  \href{http://dx.doi.org/10.1007/JHEP11(2017)094}{\emph{JHEP} {\bf 11} (2017)
  094}, [\href{https://arxiv.org/abs/1707.07685}{{\tt 1707.07685}}].

\bibitem{Chala:2012af}
M.~Chala, \emph{{$h \rightarrow \gamma\gamma$ excess and Dark Matter from
  Composite Higgs Models}},
  \href{http://dx.doi.org/10.1007/JHEP01(2013)122}{\emph{JHEP} {\bf 01} (2013)
  122}, [\href{https://arxiv.org/abs/1210.6208}{{\tt 1210.6208}}].

\bibitem{Ballesteros:2017xeg}
G.~Ballesteros, A.~Carmona and M.~Chala, \emph{{Exceptional Composite Dark
  Matter}},  \href{https://arxiv.org/abs/1704.07388}{{\tt 1704.07388}}.

\bibitem{Gripaios:2016mmi}
B.~Gripaios, M.~Nardecchia and T.~You, \emph{{On the Structure of Anomalous
  Composite Higgs Models}},
  \href{http://dx.doi.org/10.1140/epjc/s10052-017-4603-5}{\emph{Eur. Phys. J.}
  {\bf C77} (2017) 28}, [\href{https://arxiv.org/abs/1605.09647}{{\tt
  1605.09647}}].

\bibitem{Giudice:2007fh}
G.~F. Giudice, C.~Grojean, A.~Pomarol and R.~Rattazzi, \emph{{The
  Strongly-Interacting Light Higgs}},
  \href{http://dx.doi.org/10.1088/1126-6708/2007/06/045}{\emph{JHEP} {\bf 06}
  (2007) 045}, [\href{https://arxiv.org/abs/hep-ph/0703164}{{\tt
  hep-ph/0703164}}].

\bibitem{Chala:2017sjk}
M.~Chala, G.~Durieux, C.~Grojean, L.~de~Lima and O.~Matsedonskyi,
  \emph{{Minimally extended SILH}},
  \href{https://arxiv.org/abs/1703.10624}{{\tt 1703.10624}}.

\bibitem{Lee:1977eg}
B.~W. Lee, C.~Quigg and H.~B. Thacker, \emph{{Weak Interactions at Very
  High-Energies: The Role of the Higgs Boson Mass}},
  \href{http://dx.doi.org/10.1103/PhysRevD.16.1519}{\emph{Phys. Rev.} {\bf D16}
  (1977) 1519}.

\bibitem{Panico:2015jxa}
G.~Panico and A.~Wulzer, \emph{{The Composite Nambu-Goldstone Higgs}},
  \href{http://dx.doi.org/10.1007/978-3-319-22617-0}{\emph{Lect. Notes Phys.}
  {\bf 913} (2016) pp.1--316}, [\href{https://arxiv.org/abs/1506.01961}{{\tt
  1506.01961}}].

\bibitem{Serra:2015xfa}
J.~Serra, \emph{{Beyond the Minimal Top Partner Decay}},
  \href{http://dx.doi.org/10.1007/JHEP09(2015)176}{\emph{JHEP} {\bf 09} (2015)
  176}, [\href{https://arxiv.org/abs/1506.05110}{{\tt 1506.05110}}].

\bibitem{Bruggisser:2016ixa}
S.~Bruggisser, F.~Riva and A.~Urbano, \emph{{Strongly Interacting Light Dark
  Matter}},  \href{https://arxiv.org/abs/1607.02474}{{\tt 1607.02474}}.

\bibitem{Balkin:2017yns}
R.~Balkin, G.~Perez and A.~Weiler, \emph{{Little composite dark matter}},
  \href{https://arxiv.org/abs/1707.09980}{{\tt 1707.09980}}.

\bibitem{Ade:2013zuv}
{\scshape Planck} collaboration, P.~A.~R. Ade et~al., \emph{{Planck 2013
  results. XVI. Cosmological parameters}},
  \href{http://dx.doi.org/10.1051/0004-6361/201321591}{\emph{Astron.
  Astrophys.} {\bf 571} (2014) A16},
  [\href{https://arxiv.org/abs/1303.5076}{{\tt 1303.5076}}].

\bibitem{Akerib:2016vxi}
{\scshape LUX} collaboration, D.~S. Akerib et~al., \emph{{Results from a search
  for dark matter in the complete LUX exposure}},
  \href{http://dx.doi.org/10.1103/PhysRevLett.118.021303}{\emph{Phys. Rev.
  Lett.} {\bf 118} (2017) 021303},
  [\href{https://arxiv.org/abs/1608.07648}{{\tt 1608.07648}}].

\bibitem{Szydagis:2016few}
{\scshape LUX, LZ} collaboration, M.~Szydagis, \emph{{The Present and Future of
  Searching for Dark Matter with LUX and LZ}},  in \emph{{38th International
  Conference on High Energy Physics (ICHEP 2016) Chicago, IL, USA, August
  03-10, 2016}}, 2016.
\newblock \href{https://arxiv.org/abs/1611.05525}{{\tt 1611.05525}}.

\bibitem{Alarcon:2011zs}
J.~M. Alarcon, J.~Martin~Camalich and J.~A. Oller, \emph{{The chiral
  representation of the $\pi N$ scattering amplitude and the pion-nucleon sigma
  term}}, \href{http://dx.doi.org/10.1103/PhysRevD.85.051503}{\emph{Phys. Rev.}
  {\bf D85} (2012) 051503}, [\href{https://arxiv.org/abs/1110.3797}{{\tt
  1110.3797}}].

\bibitem{Alarcon:2012nr}
J.~M. Alarcon, L.~S. Geng, J.~Martin~Camalich and J.~A. Oller, \emph{{The
  strangeness content of the nucleon from effective field theory and
  phenomenology}},
  \href{http://dx.doi.org/10.1016/j.physletb.2014.01.065}{\emph{Phys. Lett.}
  {\bf B730} (2014) 342--346}, [\href{https://arxiv.org/abs/1209.2870}{{\tt
  1209.2870}}].

\bibitem{Duerr:2015aka}
M.~Duerr, P.~Fileviez~P{\'e}rez and J.~Smirnov, \emph{{Scalar Dark Matter:
  Direct vs. Indirect Detection}},
  \href{http://dx.doi.org/10.1007/JHEP06(2016)152}{\emph{JHEP} {\bf 06} (2016)
  152}, [\href{https://arxiv.org/abs/1509.04282}{{\tt 1509.04282}}].

\bibitem{Aad:2015kqa}
{\scshape ATLAS} collaboration, G.~Aad et~al., \emph{{Search for production of
  vector-like quark pairs and of four top quarks in the lepton-plus-jets final
  state in $pp$ collisions at $\sqrt{s}=8$ TeV with the ATLAS detector}},
  \href{http://dx.doi.org/10.1007/JHEP08(2015)105}{\emph{JHEP} {\bf 08} (2015)
  105}, [\href{https://arxiv.org/abs/1505.04306}{{\tt 1505.04306}}].

\bibitem{ATLAS-CONF-2016-102}
{\scshape ATLAS} collaboration, \emph{{Search for pair production of heavy
  vector-like quarks decaying to high-$p_T$ $W$ bosons and b quarks in the
  lepton-plus-jets final state in pp collisions at $\sqrt{s}$=13 TeV with the
  ATLAS detector}},  Tech. Rep. ATLAS-CONF-2016-102, CERN, Geneva, Sep, 2016.

\bibitem{ATLAS-CONF-2016-104}
{\scshape ATLAS} collaboration, \emph{{Search for new phenomena in $t\bar{t}$
  final states with additional heavy-flavour jets in $pp$ collisions at
  $\sqrt{s}=13$ TeV with the ATLAS detector}},  Tech. Rep. ATLAS-CONF-2016-104,
  CERN, Geneva, Sep, 2016.

\bibitem{ATLAS-CONF-2017-015}
{\scshape ATLAS} collaboration, \emph{{Search for pair production of
  vector-like top quarks in events with one lepton and an invisibly decaying Z
  boson in $\sqrt{s} = 13$ TeV pp collisions at the ATLAS detector}},  Tech.
  Rep. ATLAS-CONF-2017-015, CERN, Geneva, Mar, 2017.

\bibitem{CMS:2016hxa}
{\scshape CMS} collaboration, \emph{{Search for direct top squark pair
  production in the fully hadronic final state in proton-proton collisions at
  sqrt(s) = 13 TeV corresponding to an integrated luminosity of 12.9/fb}},
  Tech. Rep. CMS-PAS-SUS-16-029, CERN, Geneva, 2016.

\bibitem{Aad:2014efa}
{\scshape ATLAS} collaboration, G.~Aad et~al., \emph{{Search for pair and
  single production of new heavy quarks that decay to a $Z$ boson and a
  third-generation quark in $pp$ collisions at $\sqrt{s}=8$ TeV with the ATLAS
  detector}}, \href{http://dx.doi.org/10.1007/JHEP11(2014)104}{\emph{JHEP} {\bf
  11} (2014) 104}, [\href{https://arxiv.org/abs/1409.5500}{{\tt 1409.5500}}].

\bibitem{Alwall:2014hca}
J.~Alwall, R.~Frederix, S.~Frixione, V.~Hirschi, F.~Maltoni, O.~Mattelaer
  et~al., \emph{{The automated computation of tree-level and next-to-leading
  order differential cross sections, and their matching to parton shower
  simulations}}, \href{http://dx.doi.org/10.1007/JHEP07(2014)079}{\emph{JHEP}
  {\bf 07} (2014) 079}, [\href{https://arxiv.org/abs/1405.0301}{{\tt
  1405.0301}}].

\bibitem{Bellm:2015jjp}
J.~Bellm et~al., \emph{{Herwig 7.0/Herwig$++$ 3.0 release note}},
  \href{http://dx.doi.org/10.1140/epjc/s10052-016-4018-8}{\emph{Eur. Phys. J.}
  {\bf C76} (2016) 196}, [\href{https://arxiv.org/abs/1512.01178}{{\tt
  1512.01178}}].

\bibitem{Bellm:2017bvx}
J.~Bellm et~al., \emph{{Herwig 7.1 Release Note}},
  \href{https://arxiv.org/abs/1705.06919}{{\tt 1705.06919}}.

\bibitem{Alloul:2013bka}
A.~Alloul, N.~D. Christensen, C.~Degrande, C.~Duhr and B.~Fuks,
  \emph{{FeynRules 2.0 - A complete toolbox for tree-level phenomenology}},
  \href{http://dx.doi.org/10.1016/j.cpc.2014.04.012}{\emph{Comput. Phys.
  Commun.} {\bf 185} (2014) 2250--2300},
  [\href{https://arxiv.org/abs/1310.1921}{{\tt 1310.1921}}].

\bibitem{Gleisberg:2008ta}
T.~Gleisberg, S.~Hoeche, F.~Krauss, M.~Schonherr, S.~Schumann, F.~Siegert
  et~al., \emph{{Event generation with SHERPA 1.1}},
  \href{http://dx.doi.org/10.1088/1126-6708/2009/02/007}{\emph{JHEP} {\bf 02}
  (2009) 007}, [\href{https://arxiv.org/abs/0811.4622}{{\tt 0811.4622}}].

\bibitem{Gleisberg:2008fv}
T.~Gleisberg and S.~Hoeche, \emph{{Comix, a new matrix element generator}},
  \href{http://dx.doi.org/10.1088/1126-6708/2008/12/039}{\emph{JHEP} {\bf 12}
  (2008) 039}, [\href{https://arxiv.org/abs/0808.3674}{{\tt 0808.3674}}].

\bibitem{Cacciari:2011ma}
M.~Cacciari, G.~P. Salam and G.~Soyez, \emph{{FastJet User Manual}},
  \href{http://dx.doi.org/10.1140/epjc/s10052-012-1896-2}{\emph{Eur. Phys. J.}
  {\bf C72} (2012) 1896}, [\href{https://arxiv.org/abs/1111.6097}{{\tt
  1111.6097}}].

\bibitem{Buckley:2010ar}
A.~Buckley, J.~Butterworth, L.~Lonnblad, D.~Grellscheid, H.~Hoeth, J.~Monk
  et~al., \emph{{Rivet user manual}},
  \href{http://dx.doi.org/10.1016/j.cpc.2013.05.021}{\emph{Comput. Phys.
  Commun.} {\bf 184} (2013) 2803--2819},
  [\href{https://arxiv.org/abs/1003.0694}{{\tt 1003.0694}}].

\bibitem{Rehermann:2010vq}
K.~Rehermann and B.~Tweedie, \emph{{Efficient Identification of Boosted
  Semileptonic Top Quarks at the LHC}},
  \href{http://dx.doi.org/10.1007/JHEP03(2011)059}{\emph{JHEP} {\bf 03} (2011)
  059}, [\href{https://arxiv.org/abs/1007.2221}{{\tt 1007.2221}}].

\bibitem{Cohen:2014hxa}
T.~Cohen, R.~T. D'Agnolo, M.~Hance, H.~K. Lou and J.~G. Wacker, \emph{{Boosting
  Stop Searches with a 100 TeV Proton Collider}},
  \href{http://dx.doi.org/10.1007/JHEP11(2014)021}{\emph{JHEP} {\bf 11} (2014)
  021}, [\href{https://arxiv.org/abs/1406.4512}{{\tt 1406.4512}}].

\bibitem{Sanz:2015sua}
V.~Sanz and J.~Setford, \emph{{Composite Higgses with seesaw EWSB}},
  \href{http://dx.doi.org/10.1007/JHEP12(2015)154}{\emph{JHEP} {\bf 12} (2015)
  154}, [\href{https://arxiv.org/abs/1508.06133}{{\tt 1508.06133}}].

\bibitem{Belanger:2010pz}
G.~Belanger, F.~Boudjema, A.~Pukhov and A.~Semenov, \emph{{micrOMEGAs: A Tool
  for dark matter studies}},
  \href{http://dx.doi.org/10.1393/ncc/i2010-10591-3}{\emph{Nuovo Cim.} {\bf
  C033N2} (2010) 111--116}, [\href{https://arxiv.org/abs/1005.4133}{{\tt
  1005.4133}}].

\bibitem{Englert:2014uua}
C.~Englert, A.~Freitas, M.~M. M{\"u}hlleitner, T.~Plehn, M.~Rauch, M.~Spira
  et~al., \emph{{Precision Measurements of Higgs Couplings: Implications for
  New Physics Scales}},
  \href{http://dx.doi.org/10.1088/0954-3899/41/11/113001}{\emph{J. Phys.} {\bf
  G41} (2014) 113001}, [\href{https://arxiv.org/abs/1403.7191}{{\tt
  1403.7191}}].

\bibitem{Fujii:2015jha}
K.~Fujii et~al., \emph{{Physics Case for the International Linear Collider}},
  \href{https://arxiv.org/abs/1506.05992}{{\tt 1506.05992}}.

\end{thebibliography}\endgroup

\end{document}